\documentclass[fleqn,usenatbib]{mnras}

\usepackage{newtxtext,newtxmath}
\usepackage[T1]{fontenc}

\DeclareRobustCommand{\VAN}[3]{#2}
\let\VANthebibliography\thebibliography
\def\thebibliography{\DeclareRobustCommand{\VAN}[3]{##3}\VANthebibliography}

\usepackage{graphicx}	% Including figure files
\usepackage{amsmath}	% Advanced maths commands
\usepackage{lipsum}
\usepackage{xcolor}

\graphicspath{{./figures/}}
\usepackage[normalem]{ulem}

\newcommand{\MZgR}{MZ$_{\rm g}$R}
\newcommand{\MZsR}{MZ$_{*}$R}

\newcommand{\diff}[2]{\frac{{\rm d} #1}{{\rm d} #2}}

\defcitealias{Torrey_2018}{T18} 
\defcitealias{Springel_Hernquist_2003}{SH03}
\defcitealias{Schaye_DallaVechhia_2008}{SDV08}
\defcitealias{Mannucci_2010}{M10}
\defcitealias{Andrews_Martini_2013}{AM13}
\defcitealias{Curti_2020}{C20}

\let\oldsout\sout
\renewcommand{\sout}[1]{{\color{red}\oldsout{#1}}}

\newif\ifedits
\editsfalse

\newif\ifsecondedit
\secondeditfalse

\newif\ifthirdedit
\thirdeditfalse

\ifedits
    \newcommand{\edit}[1]{{\color{red} #1}}
\else
    \newcommand{\edit}[1]{#1}
\fi
\ifsecondedit
    \newcommand{\secondedit}[1]{{\color{red} #1}}
\else
    \newcommand{\secondedit}[1]{#1}
\fi
\ifthirdedit
    \newcommand{\thirdedit}[1]{{\color{red} #1}}
\else
    \newcommand{\thirdedit}[1]{#1}
\fi
\title[Interplay of Stellar and Gas-Phase Metallicities]{Interplay of Stellar and Gas-Phase Metallicities: Unveiling Insights for Stellar Feedback Modeling with Illustris, IllustrisTNG, and EAGLE}

\author[Garcia et al.]{Alex M. Garcia$^{1,2}$\thanks{E-mail: alexgarcia@virginia.edu},
Paul Torrey$^{1,2}$, 
Kathryn Grasha$^{3,5,6}$\thanks{ARC DECRA Fellow}, %
Lars Hernquist$^{4}$,
Sara Ellison$^{7}$,\newauthor
Henry R.M. Zovaro$^{3,5}$,
Z.S. Hemler$^{8}$,
Erica J. Nelson$^{9}$,
Lisa J. Kewley$^{4}$
\\
$^{1}$Department of Astronomy, University of Virginia, Charlottesville, VA 22904, USA\\
$^{2}$Department of Astronomy, University of Florida, 211 Bryant Space Sciences Center, Gainesville, FL 32611, USA \\
$^{3}$Research School of Astronomy \& Astrophysics, Australian National University, Canberra, Australia, 2611 \\
$^{4}$Institute for Theory and Computation, Harvard-Smithsonian Center for Astrophysics, Cambridge, MA 02138, USA \\
$^{5}$ARC Centre of Excellence for All Sky Astrophysics in 3 Dimensions (ASTRO 3D), Australia\\
$^{6}$Visiting Fellow, Harvard-Smithsonian Center for Astrophysics, 60 Garden Street, Cambridge, MA 02138, USA\\
$^{7}$Department of Physics \& Astronomy, University of Victoria, Finnerty Road, Victoria, British Columbia, V8P 1A1, Canada\\
$^{8}$Department of Astrophysical Sciences, Princeton University, Peyton Hall, Princeton, NJ, 08544, USA \\
$^{9}$Department for Astrophysical and Planetary Science, University of Colorado, Boulder, CO 80309, USA\\
}

\date{Accepted XXX. Received YYY; in original form ZZZ}

\pubyear{2023}

\begin{document}
\label{firstpage}
\pagerange{\pageref{firstpage}--\pageref{lastpage}}
\maketitle

\begin{abstract}
The metal content of galaxies provides a window into their formation in the full context of the cosmic baryon cycle. 
In this study, we examine the relationship between stellar mass and stellar metallicity (\MZsR{}) in the hydrodynamic simulations Illustris, TNG, and EAGLE to understand the global properties of stellar metallicities within the feedback paradigm employed by these simulations.
Interestingly, we observe significant variations in the overall normalization and redshift evolution of the \MZsR{} across the three simulations.
However, all simulations consistently demonstrate a tertiary dependence on the specific star formation rate (sSFR) of galaxies.
This finding parallels the relationship seen in both simulations and observations between stellar mass, gas-phase metallicity, and some proxy of galaxy gas content (e.g., SFR, gas fraction, atomic gas mass).
Since we find this correlation exists in all three simulations, each employing a sub-grid treatment of the dense, star-forming interstellar medium (ISM) to simulate smooth stellar feedback, we interpret this result as a \edit{fairly} general feature of simulations of this kind.
Furthermore, with a toy analytic model, we propose that the tertiary correlation in the stellar component is sensitive to the extent of the ``burstiness'' of feedback within galaxies. 
\end{abstract}


%%%%%%%%%%%%%%%%%%%%%%%%%%%%%%%%%%%%%%%%%%%%%%%%%%

%%%%%%%%%%%%%%%%% BODY OF PAPER %%%%%%%%%%%%%%%%%%

\section{Introduction}
\label{sec:intro}

Heavy elements are \secondedit{valuable tracers} of the gas flows that are an integral part of galaxy formation.
As aging stellar populations return metals to the interstellar medium (ISM), these metals are then dispersed and advected with any subsequent gas flows, and are incorporated into future generations of stars \citep[][]{Kobayashi_2020}.
Thus, the metal content of galaxies is a balance that is sensitive not only to the amount of heavy elements produced, but also on the efficacy of gas mixing within the ISM \citep[e.g.,][]{Elmegreen_1999,Veilleux_2005}, prevalence of outflows \citep[e.g.,][]{Veilleux_2020}, and rate of fresh/pristine gas accretion \citep[][]{Keres_2005}.
Since the metals are intrinsically tied to these physical processes within a galaxy, they may act as an observable diagnostic for the physics driving galaxy evolution \citep[][]{Dalcanton_2007,Kewley_2019}.
For example, since outflows depend strongly on stellar feedback within a galaxy and outflows have a distinct impact on galactic metallicity, the metal content of galaxies can thus be used as a probe of stellar feedback.

\edit{On large scales}, trends between physical properties and metallicity within galaxy populations have been observed.
For example, the \secondedit{stellar mass-gas-phase metallicity} relation (\MZgR) describes a trend of increasing metallicity with increasing galaxy mass \citep[][]{Tremonti_2004,Lee_2006} and is regarded as a key scaling relation between the stellar masses of galaxies and the metallicity of the gas component.
% At all redshifts, high mass galaxies are observed to be more metal enriched compared to lower mass systems \red{references}.
The reason that high mass galaxies correspond to elevated metallicities has been postulated to be a result of the high mass galaxies having either higher effective yields (i.e. retaining a higher fraction of their produced metals in the gas-phase; \citeauthor{Tremonti_2004}~\citeyear{Tremonti_2004}), lower gas fractions (which results in higher metallicities at a fixed amount of metal mass; \citeauthor{Pasquali_2012}~\citeyear{Pasquali_2012}), or perhaps both.
The slope of the \MZgR{} has been noted to change as a function of mass \edit{with a steep power-law relation at low and intermediate masses that flattens at the highest masses} \citep[e.g.,][]{Blanc_2019}.
\edit{These changes in slope are} attributed to the efficiency (or lack thereof) of a number of different physical processes, including (but not limited to) stellar feedback, ISM turbulence, and the mixing of enriched and pristine gas in the circumgalactic medium (CGM).
Moreover, there is a marked evolution in the {\MZgR} with time, whereby higher redshift galaxies form a clear {\MZgR} but with a lower overall normalisation \citep[e.g.,][]{Savaglio_2005,Maiolino_2008,Zahid_2011,Langeroodi_2022}.
The redshift evolution is attributed to the lack of chemical maturity at high redshift wherein fewer generations of stars have formed to synthesize heavier elements.
% This redshift evolution can be attributed to larger gas fractions in high redshift systems.
The {\MZgR} itself helps to reveal the net trends that must be present in terms of the gas mass evolution -- coupled to the outflow properties -- with galaxy mass and redshift for the metal content of galaxies to appropriately evolve.

In addition to the {\MZgR}, there is also a stellar mass-stellar metallicity relation (\MZsR) whereby the stellar metallicity increases with increasing stellar mass much in the same way as the {\MZgR} \citep[e.g.][]{Gallazzi_2005,Panter_2008}. 
This relationship, much like its gas-phase counterpart, follows a trend of increasing metals with increasing stellar mass before flattening at high masses \citep[e.g.,][]{Sommariva_2012,Zahid_2017}.
Just as the overall normalisation of the \MZgR{} decreases as a function of redshift, so too does the normalisation of the \MZsR{} \citep[e.g.,][]{Gallazzi_2014,Cullen_2019}.

\edit{
Many studies tend to focus on just the gas-phase or stellar metallicities, yet recent works \citep[][etc]{Topping_2020, Cullen_2021, Fraser-McKelvia_2022, Greener_2022} have shown that there is potentially a large amount of power in examining them in conjunction.
}
Physically, the link between stellar and gas-phase metals follows from the idea that newly formed stellar populations simply inherit the metallicity from the gas from which they were formed.
Thus, the stellar metallicity of the galaxy reflects the mass (or luminosity) weighted average metallicity integrated over the galaxy's formation history.
Conversely, the gas-phase traces recent evolutionary processes.
As such, the stellar metallicity is not {directly} influenced by gas inflows and/or metal production from aging stellar populations, but is instead constantly producing new stellar populations that reflect the gas phase metallicity value at the time of formation.
\edit{
Stellar metallicities tend to be systematically lower than their gas-phase counterparts across stellar mass \citep[e.g.,][]{Gallazzi_2005,Lian_2018}. 
In more detail, \cite{Gonzalez_Delgado_2014} find that the largest divergence of the two metallicities is in the lowest mass galaxies. 
\secondedit{\cite{Fraser-McKelvia_2022} attribute this divergence to the efficacy of inflows and outflows at different masses as well as a strong dependence on a galaxy's star formation history} (\citeauthor{Lian_2018} \citeyear{Lian_2018} indicate that initial mass function variations are another possibility).
\secondedit{It should be noted that the absolute values of gas-phase metallicites are sensitive to metallicity diagnostic \citep[][]{Kewley_Ellison_2008}.
Therefore trends between gas-phase and stellar metallicities are likely also sensitive to the chosen diagnostic.}
The magnitude of the discrepancy between the two metallicities \secondedit{in simulations has been seen to be both larger (\citeauthor{Yates_2012} \citeyear{Yates_2012}) and smaller (\citeauthor{DeRossi_2017} \citeyear{DeRossi_2017}) than observations.}}
\secondedit{Comparisons between simulated metallicities \thirdedit{and} observed metallicities is not always straightforward, however, without the use of mock observational techniques \citep[i.e., radiative transfer codes to generate spectra; see discussion in][]{Nelson_2018}.}

Very high redshift ($z\sim8$) measurements of gas-phase metallicities have already been attained \citep[e.g.,][]{Langeroodi_2022,Shapley_2023,Sanders_2023}, but stellar metallicities at these high redshifts are significantly more difficult to measure.
The difference between the viability \edit{of obtaining} gas-phase and stellar metallicities at high $z$ can be attributed to the different methods of deriving the two metallicities.
Briefly, gas-phase metallicities are derived from emission diagnostics, more specifically the direct and strong line methods \citep[see][for complete reviews]{Kewley_Ellison_2008,Kewley_2019}.
Stellar metallicities, on the other hand, are typically determined by absorption features in the stellar continuum.
The stellar diagnostics require high continuum signal-to-noise and as such, at present, the majority of stellar metallicity measurements come from low redshift systems \citep[e.g.,][]{Chisholm_2019}, stacking galaxies \citep[e.g.,][]{Halliday_2008,Cullen_2019,Sextl_2023}, observing lensed galaxies \citep[e.g.,][]{Patricio_2016}, or very long exposure times \citep[][]{Kriek_2019}.

It is now recognized that the scatter of galaxies about the mass-metallicity relations is not random.
Instead, systems with metallicities above (below) the {\MZgR} tend to have low (high) gas fractions and low (high) star formation rates~\citep[][etc]{Ellison_2008, Mannucci_2010, Lara_Lopez_2010, Bothwell_2013, Bothwell_2016, Yang_2022}.
This ``correlated scatter'' is naturally driven by the relationship between gas inflows, star formation, and metal production (though this is only true on the global scale; \citeauthor{Baker_Maiolino_2023} \citeyear{Baker_Maiolino_2023}).
More specifically, gas inflows tend to drive down galactic metallicities while driving up gas content and star formation rates.
Conversely, galaxies with low gas inflow rates steadily consume their gas, driving up metallicities while the gas fractions and star formation rates decay (\citeauthor{Torrey_2018}~\citeyear{Torrey_2018}; hereafter \citetalias{Torrey_2018}).
This correlated scatter can therefore be thought of as a three-parameter relation between stellar mass, gas-phase metallicity, and some proxy for the gas richness of the galaxy (gas mass, SFR, etc.).
Perhaps even more remarkably, the same three parameter fit that relates stellar mass, gas-phase metallicity, and gas richness can simultaneously fit galaxies at low- and high-redshift \edit{(\citeauthor{Cresci_2019} \citeyear{Cresci_2019}; \citeauthor{Huang_2019} \citeyear{Huang_2019}, however recent JWST observations suggest that galaxies at $z\gtrsim6$ deviate from the three parameter fit of lower redshift systems, see \citeauthor{Curti_2023} \citeyear{Curti_2023})}.
This so-called Fundamental Metallicity Relation (FMR) hints that the redshift evolution of the {\MZgR} is not disjoint from its own scatter.
Instead, the same physical processes (e.g., inflows, gas consumption) that drive galaxies about the {\MZgR} at any given redshift, also cause an evolution in the {\MZgR} with redshift.

However, since the stellar metallicities, and therefore the \MZsR{}, are not {directly} influenced by gas inflows,
% and/or newly produced metals (which enrich the gas, not the stars), 
it is not immediately obvious that there should be an equivalent FMR for stars.
Yet, observationally \edit{\citep[e.g.,][]{Zahid_2017,Sextl_2023} and in simulations \citep[][]{DeRossi_2018}}, there is evidence for this, insofar as a relationship between the \MZsR{} and star formation exists.
As of yet, there have been no observational studies systematically studying or quantifying the existence of an ``FMR for stars''.
In simulations, however, there is some evidence for this relationship not being ``fundamental'' as redshift evolution has been seen \citep[][]{Fontanot_2021}.

However, within simulations \edit{as a whole}, significant uncertainty still exists in the preferred method to model both star formation and outflows in galaxy formation models -- on which the metallicity critically depends. 
While empirical relations such as the Kennicutt-Schmidt relation~(\citeauthor{Schmidt_1959}~\citeyear{Schmidt_1959}; \citeauthor{Kennicutt_1998}~\citeyear{Kennicutt_1998}; KS relation) provide guidance or constraints for setting star formation rates (SFRs) based on the current gas content within galaxies, the implications for galactic star formation rates in very low mass or high redshift galaxies remains unclear.
Indeed, many cosmological hydrodynamic simulations (e.g., Illustris, TNG, EAGLE) employ sub-grid ISM pressurization that manifestly enforce the KS relation \citep[][]{Vogelsberger_2014b,Pillepich_2018b,Crain_2015}, giving rise to SFRs that evolve reasonably smoothly out to the highest redshifts.
Yet, models that attempt to treat both the multi-phase nature of the ISM and locality of star formation and feedback generally predict SFRs more explicitly with a significantly increased level of time variability.
\citep[i.e., burstiness;][]{Faucher-Giguere_2018}.

Notably, both ``smooth'' and ``bursty'' galaxy formation models are able to match a wide range of galaxy scaling relations (e.g., the galaxy stellar mass function, star formation main sequence, \thirdedit{mass-metallicity} relation; see, e.g., \citeauthor{Hopkins_2014} \citeyear{Hopkins_2014}; \citeauthor{Vogelsberger_2014a} \citeyear{Vogelsberger_2014a}).
Yet, the burstiness or time-variability of star formation (and its associated feedback) is of somewhat critical importance to our understanding of galaxy formation.
In particular, bursty feedback -- and the rapid, periodic, expulsion of material from the galactic center -- has been invoked as a preferred mechanism that \edit{converts} dark matter cuspy profiles into cored profiles -- alleviating an important tension between $\Lambda$CDM and observations \citep[e.g.,][]{Lazar_2020}.

In this paper, we examine this relationship between the \MZsR{} and star formation using the Illustris, IllustrisTNG, and EAGLE hydrodynamical galaxy formation simulations.
All three of these simulations are sufficiently large to capture galaxy populations where not only can the {\MZsR} be defined, but the scatter can also be characterized. 
Notably, while these simulations share several similarities in how the ISM, star formation, and metal enrichment are treated, they also model the feedback processes that shape both the {\MZgR} and {\MZsR} in detail differently.
As such, these simulations can be used to gain insight into variations resulting from changes in adopted galaxy formation models.
Moreover, we develop an analytical model to probe not only how and why the residual correlation emerges, but also how its properties change with redshift.

The structure of this paper is as follows.  
In Section~\ref{sec:methods} we describe our methods including a brief description of the simulations employed, and our definitions of stellar metallicity.
In Section~\ref{sec:results} we present our results including a comparison of the \MZsR's in Illustris, IllustrisTNG, and EAGLE, the residual correlations in the scatter, and its origin.
In Section~\ref{sec:discussion} we discuss our results, including a presentation of a simple toy model that can be used to capture the essential properties of the residual correlations, including the redshift evolution.
Finally, in Section~\ref{sec:conclusions} we provide conclusions.

\section{Methods}
\label{sec:methods}

\begin{table}
    \centering
    \begin{tabular}{lccc}
        \hline
        & {Illustris} & {TNG} & {EAGLE}\\\hline\hline
        Box side-length [Mpc] & 106.5 & 110.7 & 100 \\
        Resolution Elements & $2\times1820^3$ & $2\times1820^3$ & $2\times1504^3$\\
        Sub-Grid Physics Model & \citetalias{Springel_Hernquist_2003} & \citetalias{Springel_Hernquist_2003} & \citetalias{Schaye_DallaVechhia_2008}\\
        $m_{\rm baryon}$ [$10^6M_\odot$] & 1.26 & 1.4 & 1.81\\
        \hline
    \end{tabular}
    \caption{Summary of relevant properties of the three simulation models used in this work (Illustris, TNG, and EAGLE). $m_{\rm baryon}$ is the \edit{initial} baryonic matter mass resolution in each simulation.}
    \label{tab:simulation_details}
\end{table}

In this work, we analyse the gas-phase and stellar metallicities in Illustris, IllustrisTNG, and EAGLE galaxies to understand their dependence on star formation and stellar mass (i.e., fundamental metallicity relations).
Importantly, as will be detailed in the following sections, these three models have a commonality: sub-grid ISM pressurisation and smooth stellar feedback.
Thus, though the detailed implementations of all the models are appreciably different, we believe that any results found in all of the models \edit{should comprise a fairly generic result of simulations with this ISM treatment; however, more testing is required within (and without, as we discuss in Section~\ref{subsubec:implications}) the sub-grid ISM paradigm to confirm this.}

In this section, we briefly describe each of the simulations and our selection criteria for galaxies within the sample analysed.
All measurements are in physical units, except for the simulation box sizes, which are reported in co-moving units.
We summarize the details of this section in Table~\ref{tab:simulation_details}.

\subsection{Simulation Details}
\label{subsec:simulations}

\subsubsection{Illustris}
\label{subsubsec:Illustris}

For our analysis, we utilize the Illustris \citep[][]{Vogelsberger_2013,Vogelsberger_2014a,Vogelsberger_2014b,Genel_2014,Torrey_2014} cosmological simulations. 
These runs were performed with the moving-mesh hydrodynamical code \textsc{arepo} \citep[][]{Springel_2010}.
The Illustris model includes several important astrophysical processes including gravity, hydrodynamics, star formation, stellar evolution, chemical enrichment, radiative cooling and heating of the ISM, stellar feedback, black hole growth, and active galactic nuclei (AGN) feedback.
Most important for the purposes of our work are self consistent implementations of star formation, stellar feedback, and chemical enrichment.
Illustris treats the dense, star forming ISM with the effective equation of state of \citeauthor{Springel_Hernquist_2003} (\citeyear{Springel_Hernquist_2003}; hereafter,
\citetalias{Springel_Hernquist_2003}).
In this model, gas beyond a threshold density ($n_{\rm H} > 0.13$ cm$^{-3}$) forms stars probabilistically assuming a \cite{Chabrier_2003} initial mass function (IMF).
Crucially, the stars that are formed adopt their metallicity from the gas from which they form.
As the stars evolve, they return both mass and metals to the ISM.
Stellar mass return and yield tables are employed to allow for the direct simulation of time-dependent mass return and heavy element enrichment.
Illustris explicitly tracks nine different chemical species (H, He, C, N, O, Ne, Mg, Si, and Fe).

The main Illustris simulation included a single volume of size (106.5 Mpc)$^3$ at three different resolutions, which are denoted as:
Illustris-1 for the highest resolution run (with $2\times1820^3$ particles) and Illustris-3 for the lowest resolution run (with $2\times455^3$ particles).
In this work, we use the highest resolution run, Illustris-1, which we will henceforth use synonymously with Illustris itself.

\subsubsection{IllustrisTNG}
\label{subsubsec:TNG}

IllustrisTNG \citep[][hereafter TNG]{Marinacci_2018,Naiman_2018,Nelson_2018,Pillepich_2018b,Springel_2018,Pillepich_2019, Nelson_2019a, Nelson_2019b} is the successor to the original Illustris simulations.
TNG includes updates to the original Illustris model \edit{in addition to} alleviating some deficiencies.
Importantly, TNG also models the dense, star forming ISM with the \citetalias{Springel_Hernquist_2003} equation of state.
As such, these two simulations are related, yet have appreciably different physical implementations \citep[for a complete list of differences see][]{Weinberger_2017,Pillepich_2018a}.
Critical for the discussion later in this work, TNG implements redshift dependent winds. 
A wind velocity floor is enforced in TNG to ensure that low mass haloes do not have unphysically large mass loading factors.
The effect of this is that winds are more efficient at suppressing low-redshift star formation and, consequently, feedback is more effective at higher redshifts.
TNG tracks the same nine elements as Illustris, but it additionally adds an ``other metals'' item as a proxy for other metals not explicitly tracked.

Whereas the original Illustris suite consisted of only one volume at several resolution levels, TNG is comprised of three different volumes, each with their own set of resolutions.
The simulations within TNG are designated by their approximate box size -- TNG50 (51.7 Mpc)$^3$, TNG100 (110.7 Mpc)$^3$, and TNG300 (302.6 Mpc)$^3$, with the different resolution levels indicated as in the Illustris suite.
Here, we employ the highest resolution run of TNG100 (TNG100-1; $2\times1820^3$ particles, hereafter simply TNG) to make a fair comparison with the original Illustris simulation.

\subsubsection{EAGLE}
\label{subsubsection:EAGLE}

Finally, we use the ``Evolution and Assembly of GaLaxies and their Environment'' \citep[EAGLE,][]{Crain_2015, Schaye_2015, McAlpine_2016} cosmological simulations.
Unlike Illustris and TNG, EAGLE employs \edit{a heavily modified version of {\sc gadget-3} \citep{Springel_2005} employing the {\sc anarchy} formulation of smoothed particle hydrodynamics (SPH; see \citeauthor{Schaye_2015} \citeyear{Schaller_2015} Appendix A}), which alleviates problems with the original version of 
SPH \citep{Sijacki_2012, Vogelsberger_2012, Keres_2012, Torrey_2012}.  
EAGLE is a full-physics simulation that includes many of the same baryonic processes with hydrodynamics (star-formation, chemical enrichment, radiative cooling and heating, etc).
The dense (unresolved) ISM is treated with a sub-grid equation of state (\citeauthor{Schaye_DallaVechhia_2008}~\citeyear{Schaye_DallaVechhia_2008}; hereafter, \citetalias{Schaye_DallaVechhia_2008}), much like that of \citetalias{Springel_Hernquist_2003}.
\edit{The \citetalias{Schaye_DallaVechhia_2008} prescription forms stars according to a \cite{Chabrier_2003} IMF from the dense ISM gas.
The density threshold for star formation is given by the metallicity-dependent transition from atomic to molecular gas computed by \cite{Schaye_2004}, \thirdedit{with an additional temperature-dependent criterion \citep[see][their Section 4.3]{Schaye_2015}}.}
% The \citetalias{Schaye_DallaVechhia_2008} prescription, allows gas above a certain density (though slightly less dense than \citetalias{Springel_Hernquist_2003}; $n_{\rm H} > 0.1~{\rm cm}^{-3}$) to form stars according to a \cite{Chabrier_2003} IMF.
Stellar populations evolve according to the \edit{\cite{Wiersma_2009b}} evolutionary model and eventually return their mass and metals back into the ISM.
While the \citetalias{Springel_Hernquist_2003} equation of state describes the multiphase ISM with the average gas density, \citetalias{Schaye_DallaVechhia_2008} use a polytropic equation of state.
% The \citetalias{Schaye_DallaVechhia_2008} prescription produces much stronger winds than that of \citetalias{Springel_Hernquist_2003}, resulting in galaxy-wide perturbations.
\edit{EAGLE tracks eleven different chemical species (H, He, C, N, O, Ne, Mg, Si, S, Ca, and Fe).}

\edit{EAGLE is comprised of several simulations ranging from size \secondedit{$(12~{\rm Mpc})^3$} to $(100~{\rm Mpc})^3$.
For this work, as an even-handed comparison to the selected Illustris and TNG runs, we employ data products \thirdedit{of} an intermediate resolution ($2\times1504^3$ particles) run with a box-size of ($100~{\rm Mpc})^3$ referred to as {\sc RefL0100N1504} (hereafter simply EAGLE).}

\subsection{Galaxy Selection}
\label{subsec:selectionCriteria}

In all of the aforementioned simulations, gravitationally-bound substructures are identified using a {\sc subfind} algorithm \citep{Springel_2001,Dolag_2009} which relies on a friends-of-friends \citep[FoF;][]{Davis_1985} algorithm to identify parent groups.
\secondedit{In this paper, the total mass identified by {\sc subfind} is used for calculating global galaxy properties whereas observational works typically estimate quantities within a given aperture.}
We limit the sample from each simulation to central galaxies with stellar masses $8.0 < \log(M_*~[M_\odot]) < 12.0$ and \secondedit{(total)} gas mass\footnote{\secondedit{We note this gas mass threshold includes both \thirdedit{the} star forming and non-star forming gas.}} limits of $\log(M_{\rm gas}~[M_\odot]) > 8.5$.
This ensures that we have $\sim10^2$ star particles and $\sim5\times10^2$ gas particles per selected galaxy, which we consider well-resolved \edit{(see Table~\ref{tab:simulation_details} for initial baryon mass resolutions of each simulation)}.
We note that while this resolution is too coarse to make statements about spatially resolved properties within the galaxies (e.g., metallicity gradients), it is sufficient for global properties of the systems.
Further, following from \cite{Donnari_2019}, \cite{Pillepich_2019}, \cite{Nelson_2021}, \cite{Hemler_2021}, and \cite{Garcia_2023}, we define a specific star formation main sequence (sSFMS) in order to select only star forming galaxies.
The sSFMS is defined through a linear-least squares fit to the median relation of galaxies with stellar mass $\log(M_*~[M_\odot]) < {10.2} $ in mass bins of {0.2 dex}.
Beyond $\log(M_*~[M_\odot]) > {10.2}$, we extrapolate the linear fit.
From this sSFMS, we define (and omit) quiescent galaxies as those whose sSFRs lie greater than 0.5 dex below this relation.
\edit{We note that our key results are not sensitive to these selection criteria for the sSFMS, which we discuss further in Appendix~\ref{appendix:sSFMS_dependence}}.

With all of these choices, \edit{at $z=0$ we obtain 44,965 galaxies in Illustris, 22,133 in TNG, and 14,720 in EAGLE.}
\edit{While all of these simulations have very similar halo mass functions, we note that the overall number of galaxies in the sample is quite different between the three. 
The difference between the number of galaxies in EAGLE and TNG can be attributed to: (i) the larger box of TNG simply containing more galaxies and (ii) \thirdedit{lower} gas fractions in EAGLE making it more sensitive to our gas mass cut at the low mass end.
Illustris, on the other hand, has an elevated stellar mass function compared to TNG, which is due to the different feedback and wind implementations \citep{Pillepich_2018a,Pillepich_2018b}, as such there are a factor of $\sim$2 more galaxies at $z=0$ than in TNG.}

\edit{Throughout all of this analysis gas-phase metallicities are from star-forming gas particles only as a fair comparison with observations, whose gas-phase metallicities are limited to emission diagnosistics from star forming regions of galaxies \citep[see][]{Kewley_Ellison_2008,Kewley_2019}.}
\secondedit{
Gas is determined to be star-forming based on criteria outlined in Section~\ref{subsubsec:Illustris} for Illustris/TNG and Section~\ref{subsubsection:EAGLE} for EAGLE.
}

We aim to offer predictions for the high redshift ($z\lesssim10$) \MZsR{} in each of these simulations.
However, due to the aforementioned mass limits of our sample, beyond $z=8$ there are not enough galaxies to create a statistically significant sample.
We therefore offer predictions for galaxies only out to $z=8$.

\section{Results}
\label{sec:results}

\begin{figure*}
    \centering
    \includegraphics[width=\linewidth]{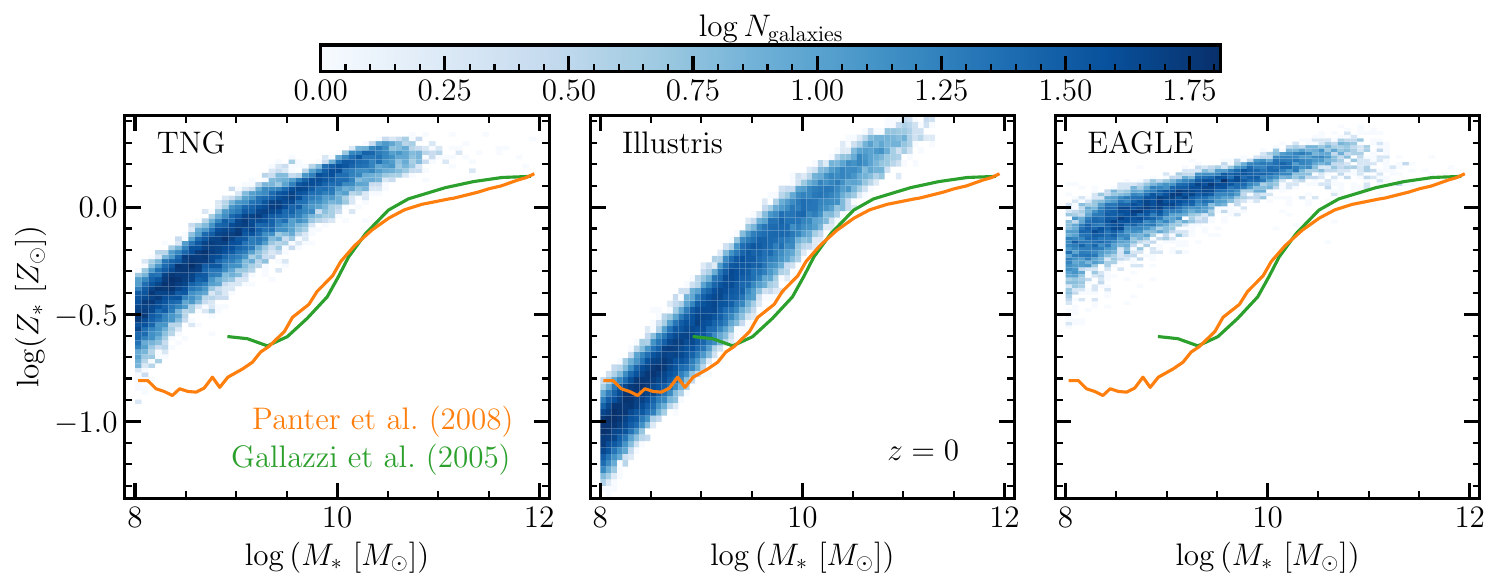}
    \caption{Stellar mass versus stellar metallicity (both scaled to solar; $Z_\odot = 0.0127$) for star forming galaxies in \edit{TNG, Illustris, and EAGLE} (left, centre, and right panels, respectively) at $z=0$. Each panel is colour-coded by the total number of galaxies within each pixel, with lighter colours corresponding to fewer galaxies.
    We overplot the \MZsR{} from local observations (green, \citeauthor{Gallazzi_2005} \citeyear{Gallazzi_2005}; \edit{orange,} \citeauthor{Panter_2008} \citeyear{Panter_2008}).
    It is important to note that comparisons between stellar metallicities from simulations to stellar metallicities from observations is not necessarily one-to-one \edit{(see text for more details)}, thus we caution the reader in any comparisons against observed \MZsR{} trends.}
    \label{fig:MZR*}
\end{figure*}

\subsection{Stellar MZR}
\label{subsec:stellar_MZR}

\begin{figure*}
    \centering
    \includegraphics[width=\linewidth]{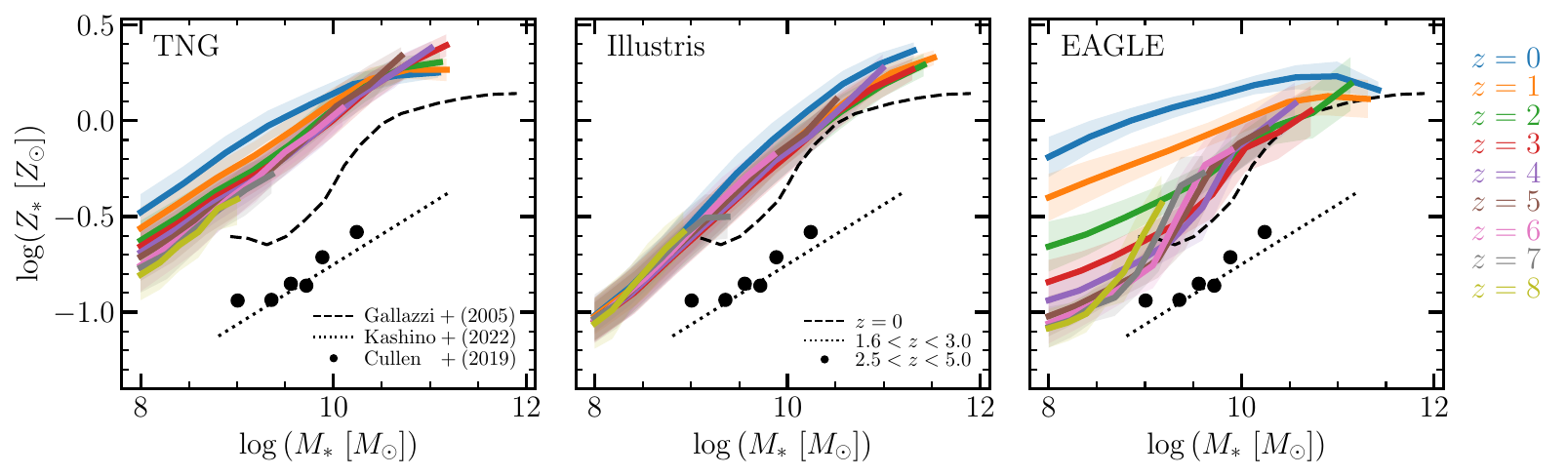}
    \caption{\MZsR{}, as defined in Section~\ref{subsec:stellar_MZR}, for star-forming galaxies in the TNG, Illustris, and EAGLE simulations (left, centre, and right panels, respectively) for $z = 0-8$. Galaxies are split into several different mass bins, and the median metallicity within each bin is measured. We require that there must be at least 10 galaxies in each mass bin shown (i.e., higher masses are less populated at higher redshift). 
    \edit{The shaded regions correspond to the standard deviation about the median mass bins. The black lines and points correspond to observations from \protect\cite{Gallazzi_2005} at $z=0$ (dashed line), \protect\cite{Kashino_2022} ranging from $1.6 < z < 3.0$ (dotted line), and \protect\cite{Cullen_2019} from $2.5 < z < 5.0$ (points).}
    }
    \label{fig:MZRallz}
\end{figure*}

{Critically, the analysis done in this work relies on the underlying relationship between the mass of a system and the total metal content locked in stars within galaxies.
To that end, we first quantify the \MZsR{} in Illustris, TNG, and EAGLE as a baseline for the rest of the results in this work.}
In these three simulations we find that a relationship between stellar mass and stellar metallicity exists, as illustrated in Figure~\ref{fig:MZR*} for each simulation at $z=0$.
We include the \MZsR{}s from \cite{Gallazzi_2005} and \cite{Panter_2008} as a point of comparison in Figure~\ref{fig:MZR*}; however, we caution the reader that comparing metallicities from observations and simulations is not even-handed \citep[see][]{Nelson_2018}.
\edit{
Specifically, computing metallicities by averaging the metal content of star particles weighted by their mass (as is done in all three simulations analysed here) produces systematically elevated stellar metallicities compared to using radiative transfer codes more akin to mock observations of galaxies \citep[][]{Guidi_2016}.
In that light, it is perhaps unsurprising that the $z=0$ \MZsR{}s presented in Figure~\ref{fig:MZR*} and higher redshift versions in Figure~\ref{fig:MZRallz} are all elevated compared to the observations.
}
\edit{Nevertheless, it should be noted that while Illustris and TNG appear to follow much closer to the \cite{Gallazzi_2005} and \cite{Panter_2008} \MZsR{}s compared to EAGLE, a higher resolution run of EAGLE with modified sub-grid parameters ({\sc recal}-L025N0752) more closely reproduces the observed \MZsR{} \citep[see][their Figure 13]{Schaye_2015}. }

In terms of the shape of the relation in the simulations, we find qualitative agreement between the three: roughly power-law with constant slope at low masses, leveling out at higher masses, qualitatively similar to the observations.
In detail, however, the three simulations exhibit different behaviours.
Specifically, we find that the \MZsR{} normalisation at $z=0$ for the lowest mass galaxies vary by nearly 1.0 dex between the simulation with the EAGLE galaxies being more enriched than those in TNG, which, in turn, are more enriched than the galaxies in Illustris.
At higher masses, $M_* \sim 10^{10} M_\odot$, however, the normalisation of the metallicities appears more similar between the different simulations. 
We also find that the slope of the power-law relation at low masses differs between the simulations: Illustris galaxies increase in metallicity more rapidly as a function of stellar mass, than those in TNG and EAGLE. 
%\red{due in part to the elevated metallicities in TNG and EAGLE (does this make sense?)}.

At $z>0$, we find that differences in normalisation persist, and, in addition, the normalisations appear to evolve differently between the various simulations.
In Figure~\ref{fig:MZRallz}, we trace the \MZsR{} back to $z=8$ for each simulation, \edit{which we compare to observations from \cite{Gallazzi_2005} at $z=0$, \cite{Kashino_2022} at $1.6 < z < 3.0$, and \cite{Cullen_2019} at $2.5 < z < 5.0$ (though, see above discussion about comparison of stellar metallicities with observations)}.
For simplicity, we define the \MZsR{} simply as the median metallicity within fixed mass bins\footnote{We note that if a particular mass bin does not have more than 10 galaxies, we omit it from our sample.}.
\edit{
The shaded regions about each \MZsR{} represents the scatter about the relation, which we define as the standard deviation of metallicities within each mass bin.
}
Consistent with the $z=0$ \MZsR{}, the higher redshift Illustris galaxies are less enriched at low masses and follow a steeper trend than their TNG and EAGLE counterparts.
However, around $z\gtrsim4$, the EAGLE galaxies have changed normalisation enough to be more consistent with Illustris and there is a significant increase in the slope of the relation.
To that end, there is a marked difference in the redshift evolution of the \MZsR{} between the three simulations: TNG has some modest normalisation evolution with lower redshift galaxies being more enriched than high redshift, EAGLE has a similar, albeit more extreme, evolution, and Illustris shows virtually no redshift evolution.
% \edit{ {\bf needs work} In particular, the evolution of the \MZsR{} is more significant than the scatter about the relation at each redshift. }
This lack of evolution in the stellar metallicities over cosmic time in Illustris has been noted previously \citep{DSouza_2018}.
These authors attribute the modest redshift evolution of the \MZsR{} shown in \cite{Gallazzi_2014} at $z\approx0.7$ as evidence that, in the absence of robust predictions at higher redshifts, the lack of redshift evolution in the \MZsR{} is physical.
However, the predictions made here in TNG and EAGLE of an \MZsR{} that evolves with redshift is in agreement with other simulation models \citep[e.g.,][]{Ma_2016,Yates_2021} and \edit{more recent} observations \citep[e.g.,][]{Choi_2014,Leethochawalit_2018,Cullen_2019,Beverage_2021,Kashino_2022}.

\subsection{Residual correlations in the scatter}
\label{subsec:residualCorrelations}

\begin{figure*}
    \centering
    \includegraphics[width=\linewidth]{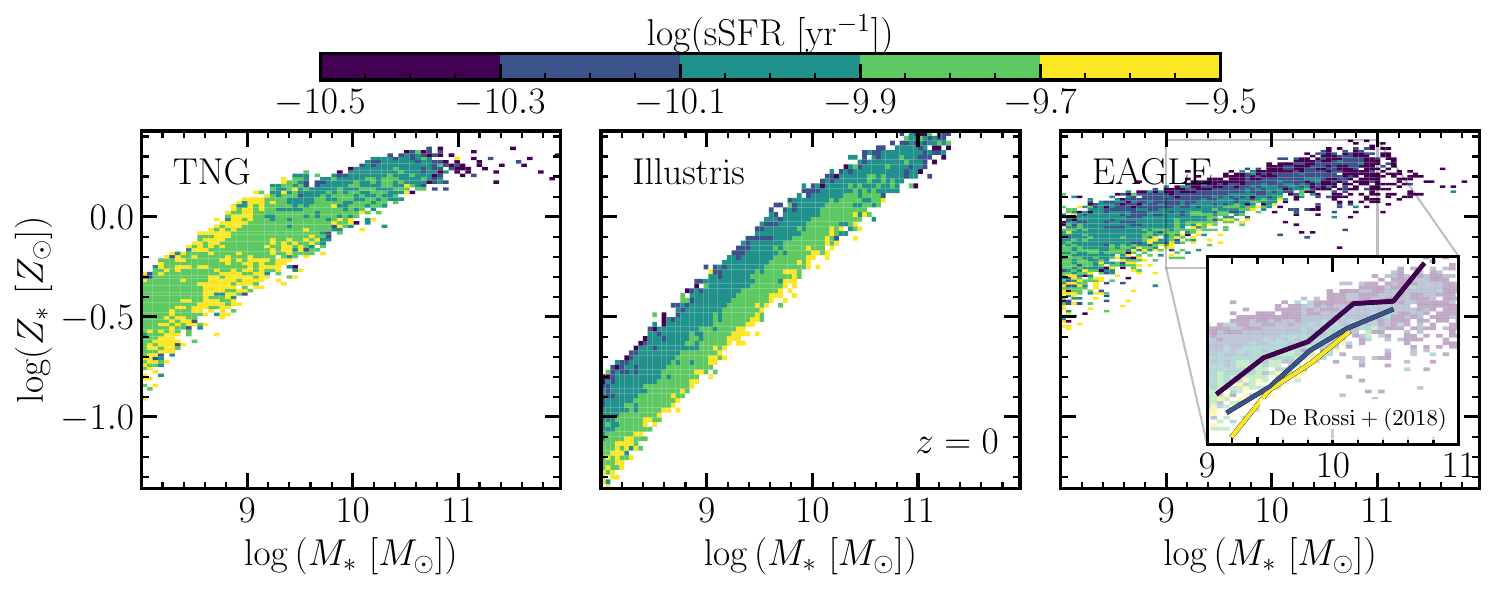}
    \caption{Same as Figure~\ref{fig:MZR*} colour coded by specific star formation rate (i.e., total star formation rate per unit stellar mass of the galaxy; sSFR). Lighter colours correspond to higher star forming galaxies while darker colours are lower star forming galaxies. At fixed stellar mass lower metallicity galaxies tend to have higher sSFRs. \edit{The inset on the right panel is a comparison with \protect\cite{DeRossi_2018} who use a higher resolution run of EAGLE at $z=0$; full discussion of this comparison can be found in Section~\ref{subsec:comparisonModels}.} Note that the left panel of TNG galaxies, at lower masses, does not clearly display this relationship at $z=0$, we discuss this further in Section~\ref{subsec:origin_MZR} and Appendix~\ref{appendix:TNGz=0}. }
    \label{fig:MZSFR*}
\end{figure*}

\begin{figure*}
    \centering
    \includegraphics[width=\linewidth]{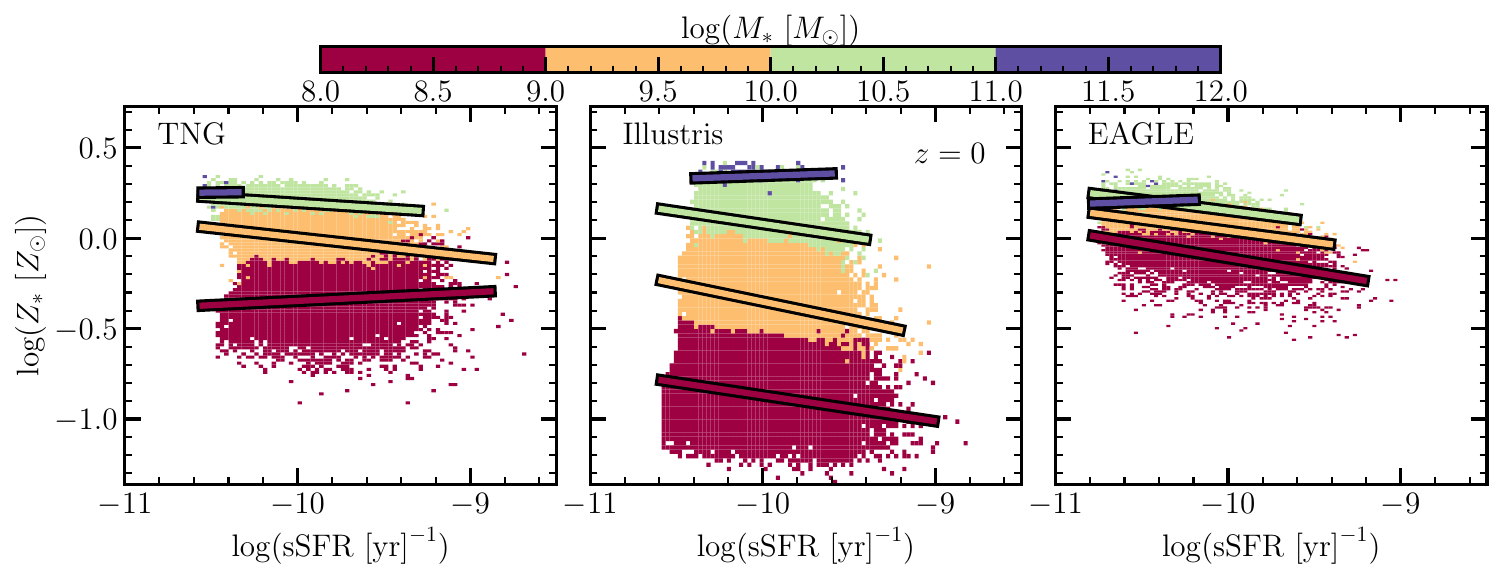}
    \caption{Stellar metallicity as a function of sSFR for star-forming galaxies in TNG, Illustris, and EAGLE (left, centre, and right, respectively) at $z=0$ colour-coded by stellar mass bins. Galaxies are in mass bins of width 1.0 dex, centered at $8.5, 9.5, 10.5, ~{\rm and}~ 11.5 \log(M_*~[M_\odot])$. Linear least-squares fits are provided by the outlined lines for each mass bin. We note that most of these slopes, across redshift, are negative, indicating a decrease in metallicity with increasing sSFR for a given mass. }
    \label{fig:MZSFR_sSFR}
\end{figure*}

Within the scatter of the \MZgR{} residual correlations have been shown to exist.
In particular, a secondary correlation with the star formation rate (SFR) or specific star formation rate (sSFR; SFR normalized by galaxy mass) has been seen both observationally \citep[e.g.,][]{Ellison_2008,Mannucci_2010,Andrews_Martini_2013} and in simulations (\edit{e.g., \citeauthor{DeRossi_2017} \citeyear{DeRossi_2017} in EAGLE;} \citetalias{Torrey_2018} in TNG).
We find that galaxies' stellar metallicities also follow this trend: at a fixed stellar mass, galaxies with lower stellar metallicities generally have higher sSFRs (Figures~\ref{fig:MZSFR*}~and~\ref{fig:MZSFR_sSFR}), \edit{echoing a similar result at $z=0$ for a higher resolution run of EAGLE by \citeauthor{DeRossi_2018} (\citeyear{DeRossi_2018}; see Section~\ref{subsec:comparisonModels} for a more in-depth comparison) }.
Figure~\ref{fig:MZSFR*} shows the \MZsR{} for each simulation at $z=0$ colour coded by their sSFR.
Qualitatively, we find that for all simulations, save for TNG at $z=0$, there is a clear residual correlation of the \MZsR{} with sSFR.
At $z=0$ in TNG, lower mass galaxies ($M_*\lesssim 10^{9.5} M_\odot$) the residual correlation appears weaker than their Illustris or EAGLE counterparts.
However, \edit{as we show in Appendix~\ref{appendix:TNGz=0}}, at $z>0$, the galaxies fall into the more familiar relation like that of Illustris and EAGLE at $z=0$.
Given the overall model similarities -- and differences -- between Illustris and TNG, we attribute the lack of a clear residual correlation at $z=0$ in TNG to the redshift-dependent wind\thirdedit{s} in the model (which was not present in Illustris).
% ; we discuss this issue further in Appendix~\ref{appendix:TNGz=0}.
% \red{Don't be dismissive of things in the Appendix. Briefly summarize here.}

More quantitatively, Figure~\ref{fig:MZSFR_sSFR} shows the stellar metallicities as a function of sSFR in fixed mass bins (coded by their colour) for each simulation at $z=0$.
The overplotted lines are the best fit within the corresponding mass bin.
We find that the majority of these best fits (at all redshifts) have negative slopes, indicating that, at a fixed stellar mass, galaxies with elevated stellar metallicities generally have lower specific star formation rates.
As previously mentioned, one notable exception to this trend is the lowest mass galaxies in TNG at $z=0$.
This complements the qualitative statements above: the slope of the correlations in TNG at $z=0$ here in the lowest mass bins ($10^{8.0}-10^{9.0}M_\odot$) is inverted.
We show in Appendix~\ref{appendix:TNGz=0} that at higher redshift, the residual correlation is much more apparent in this mass range for TNG.
Another notable exception is in the highest mass bins.
In all three simulations, the highest mass bins (purple in Figure~\ref{fig:MZSFR_sSFR}), appear much flatter than, or inverted with respect to, their lower-mass counterparts \edit{out to $z=2$ for EAGLE and $z=1$ for Illustris and TNG}.
\edit{Interestingly,} this flattening and inversion of the residual correlation within higher stellar masses has been noted previously in both simulations (\citeauthor{Yates_2012} \citeyear{Yates_2012}, {\sc l-galaxies}; \citeauthor{DeRossi_2017} \citeyear{DeRossi_2017}, EAGLE) and observations (\citeauthor{Yates_2012} \citeyear{Yates_2012}, Sloan Digitial Sky Survey DR7) \secondedit{within the gas-phase metallicities and in simulations (\citeauthor{DeRossi_2018} \citeyear{DeRossi_2018}, EAGLE) in the stellar metallicities. 
}
\edit{
\cite{DeRossi_2017,DeRossi_2018} show, using modified AGN parameters within EAGLE, that this inversion is due to AGN feedback, which is almost certainly the case in the highest mass galaxies analysed here.
}

\subsubsection{Projection of least scatter}
\label{subsubsec:leastScatter}

\begin{figure}
    \centering
    \includegraphics[width=\linewidth]{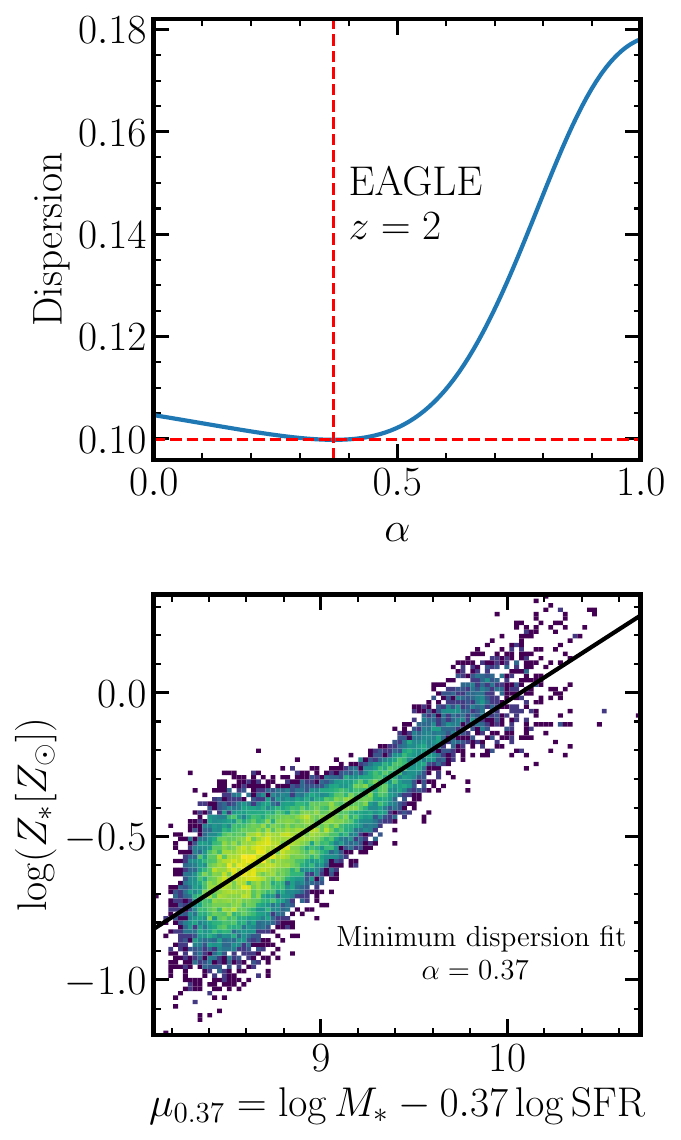}
    \caption{Demonstration of parameterisation of $M_*-Z_*-{\rm SFR}$ correlation in terms of $\mu_{\alpha}$ (Equation~\ref{eqn:mu}) following from \citetalias{Mannucci_2010}. {\bf Top:} The dispersion (i.e., standard deviation of the residuals) from the linear fit as a function of $\alpha$, the free parameter in $\mu_{\alpha}$. Shown here is the dispersions for fits of the EAGLE $z=2$ $M_*-Z_*-{\rm SFR}$ relation. The red dashed lines indicate where the dispersion is minimized and at what value of $\alpha$. The gray dashed lines represent the uncertainty in $\alpha$ values corresponding to 1\% deviations from the minimum dispersion. {\bf Bottom:} The $M_*-Z_*-{\rm SFR}$ relation parameterised by the $\mu_{\alpha}$ that minimizes the dispersion about the linear relation. For EAGLE at $z=2$, we find that the best $\alpha = 0.37$.}
    \label{fig:alpha_getter}
\end{figure}

\citeauthor{Mannucci_2010} (\citeyear{Mannucci_2010}; henceforth \citetalias{Mannucci_2010}) posit that if the metallicity, SFR, and stellar mass are all correlated, then perhaps there exists a tighter parameterisation than a traditional MZR.
They fit the mass-(gas-phase) metallicity-SFR ($M_*-Z_{\rm gas}-{\rm SFR}$) relation with a linear combination of the stellar mass and star formation rates,
\begin{equation}
    \label{eqn:mu}
    \mu_\alpha = \log M_* - \alpha\log({\rm SFR})~,
\end{equation}
where $\alpha$ is a free parameter\footnote{We note that more complex relationships parameterising this residual correlation are available \citep[e.g.,][]{Lara_Lopez_2010,Lara_Lopez_2013,Curti_2020}. We opt to use this specific parameterisation in view of its simplicity.}.
This free parameter encodes the relative importances of stellar mass and SFR.
For example, taking $\alpha=0.0$, the original form of the MZR is recovered, which would suggest that the SFR is unimportant in minimzing the scatter of the relation.
Conversely, with $\alpha=1.0$ the primary driving factor of the scatter is the {\it specific} star-formation rates (sSFR) of the galaxies.
We employ this same parameterisation for the stellar mass-stellar metallicity-SFR ($M_*-Z_*-{\rm SFR}$) relation, later in this section.

To determine the best $\mu_\alpha$ we fit a linear regression of $\log (Z_*~[Z_\odot])$ versus $\mu_\alpha$ for values of $\alpha$ ranging from 0 to 1.
Whichever value of $\alpha$ corresponds to the smallest standard deviation about fixed $\mu_{\alpha}$ bins from the linear fit is deemed the best fit (demonstrated in Figure~\ref{fig:alpha_getter} for EAGLE at $z=2$).
Further, we define an uncertainty of this parameter by taking any $\alpha$ that only varies by 1\% of the minimum dispersion (see dashed gray lines in top panel of Figure~\ref{fig:alpha_getter}).

\begin{figure*}
    \centering
    \includegraphics[width=\linewidth]{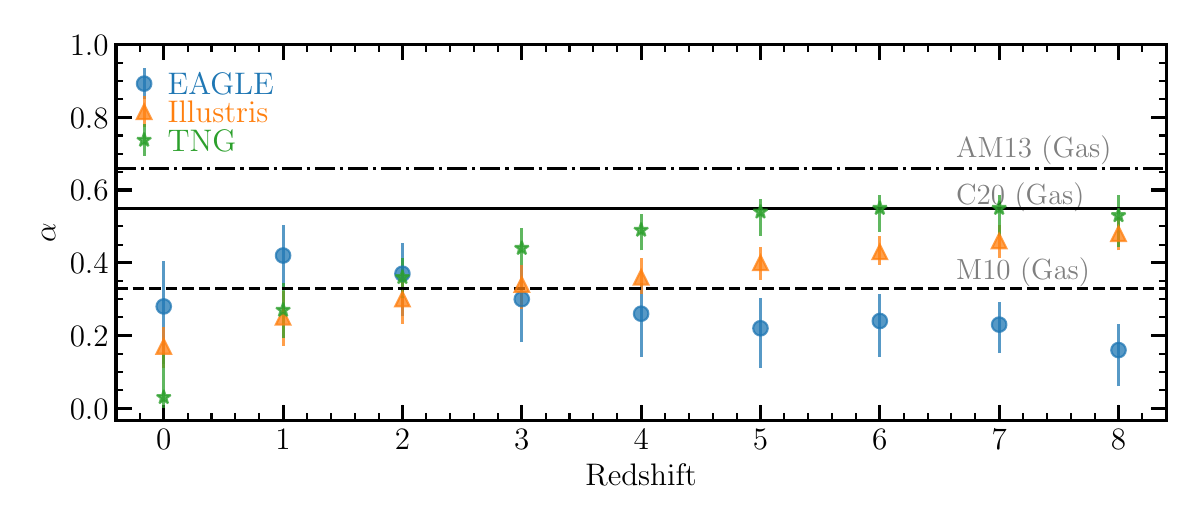}
    \caption{$\alpha$, the relative weighting of star formation to stellar mass in the three component relation ($M_*-Z_*-{\rm SFR}$; as defined in Section~\ref{subsubsec:leastScatter}), versus redshift in EAGLE (blue \edit{cirlces}), Illustris (orange \edit{triangles}), and TNG (green \edit{stars}). The dashed line represents the relative importance found by \citetalias{Mannucci_2010} ($\alpha=0.33$), the dot-dashed line is that found by \citetalias{Andrews_Martini_2013} ($\alpha=0.66$), and the solid line is \citetalias{Curti_2020} ($\alpha=0.55$; note that all these works present $\alpha$ with the gas-phase metals). The errorbars correspond to the uncertainty associated with 1\% deviations from the minimum dispersion of the best fit $\mu_{\alpha}$ (see Figure~\ref{fig:alpha_getter} and Section~\ref{subsubsec:leastScatter} for more details). }
    \label{fig:alpha_redshift}
\end{figure*}

Figure~\ref{fig:alpha_redshift} shows the evolution of the $\alpha$ parameter as a function of redshift in each of the simulations with the errorbars corresponding to the aforementioned 1\% uncertainty of $\alpha$.
We compare against observational values determined in \citetalias{Mannucci_2010} ($\alpha=0.33$), \citeauthor{Andrews_Martini_2013} (\citeyear{Andrews_Martini_2013}, henceforth \citetalias{Andrews_Martini_2013}; $\alpha=0.66$), and \citeauthor{Curti_2020} (\citeyear{Curti_2020}, henceforth \citetalias{Curti_2020}; $\alpha=0.55$) for the gas-phase.
Since these observational values specifically refer to the gas-phase and not stellar metallicities, we caution the reader to take comparisons against observations lightly.
Another important consideration is that the relative weighting of the mass and SFR has been shown to depend strongly on the metallicity diagnostic used (\citetalias{Andrews_Martini_2013}).

Regardless, we ultimately find a non-zero evolution in the best fit $\alpha$ as a function of redshift.
At $z=0$, we find relatively weak dependence on the SFR; in fact, in TNG there is virtually no dependence on the SFR.
This lack of importance placed on the SFR is perhaps unsurprising in light of the lack of a significant $M_*-Z_*-{\rm SFR}$ relation at that redshift in TNG (mentioned in the previous section and discussed further in Appendix~\ref{appendix:TNGz=0}).
In Illustris and TNG, $\alpha$ increases slightly with increasing redshift.
In Eagle, on the other hand, $\alpha$ decreases weakly with increasing redshift.
While the $M_*-Z_{\rm gas}-{\rm SFR}$ relation is oftentimes referred to as the fundamental metallicity relation (FMR; \citetalias{Mannucci_2010}; \citetalias{Andrews_Martini_2013}; \citetalias{Curti_2020}) and has been shown to have no evolution in the relation out to $z\sim2.5-3.5$ (e.g., \citetalias{Mannucci_2010}; \citeauthor{Lara_Lopez_2010}~\citeyear{Lara_Lopez_2010}), we avoid interpreting the $M_*-Z_*-{\rm SFR}$ as an ``FMR for stars'' due to the  redshift evolution seen in each of the simulations presented here.
Interestingly, the redshift dependence on this relationship is also noted in \cite{Fontanot_2021} with a semi-analytic model (SAM; see Section~\ref{subsec:comparisonModels} for a more in-depth comparison with this model).
Evidently, as we have shown, this $M_*-Z_*-{\rm SFR}$ relationship is not ``fundamental'' insofar as it evolves with time.

\subsection{Origin of \texorpdfstring{$M_*-Z_*-{\rm SFR}$}{M-Z-SFR} Relation}
\label{subsec:origin_MZR}

\begin{figure*}
    \centering
    \includegraphics[width=\linewidth]{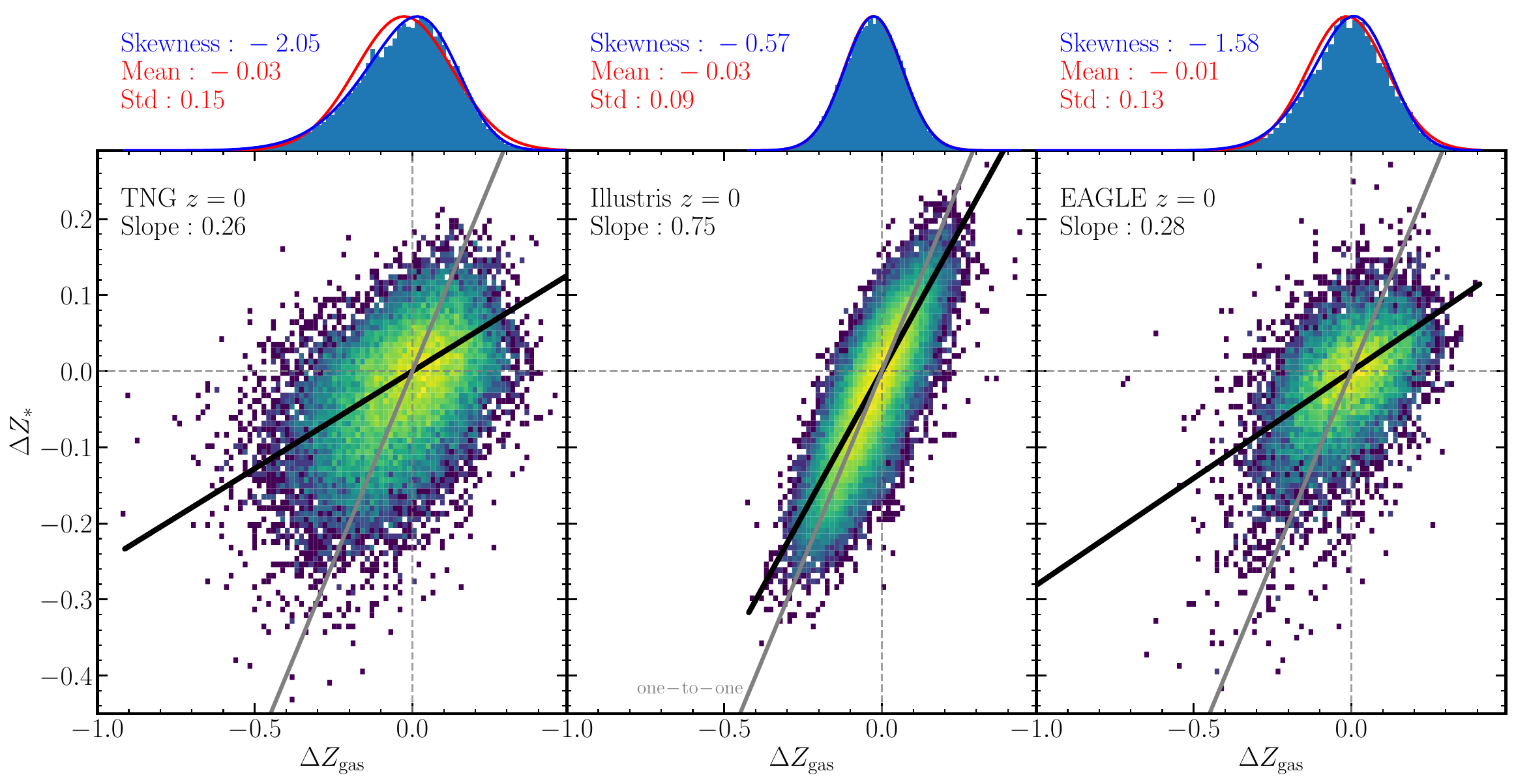}
    \caption{{\bf Bottom Panels:} Offsets of both the gas-phase (abscissa) and stellar (ordinate) metallicities of star forming galaxies in EAGLE, Illustris, and TNG (left to right) at $z=0$ from the MZRs. In each panel, a linear least squares best fit line to this relation is overplotted (black line; fixing the intercept to the origin) as well as a one-to-one line (gray solid line). The slope of the linear least squares fit is the relevant quantity in this relation and allows us to quantify how correlated the two metallicities are to each other; a slope of unity would suggest that the gas-phase and stellar metals are exactly correlated, while a slope of 0 suggests that they are uncorrelated. {\bf Top Panels:} Normalised distributions of the offsets from each simulation's \MZgR{}. We fit both a Gaussian distribution (red) and skewed Gaussian (blue) to the distribution. The typical mean and standard deviation of these distributions (across redshift) is $\sim$0.0 dex and $\sim$0.12 dex, respectively. We find that the distributions have some non-zero skew, which we discuss in more detail in Section~\ref{subsec:analyticModel}.}
    \label{fig:offset_relations}
\end{figure*}

\begin{figure*}
    \centering
    \includegraphics[width=\linewidth]{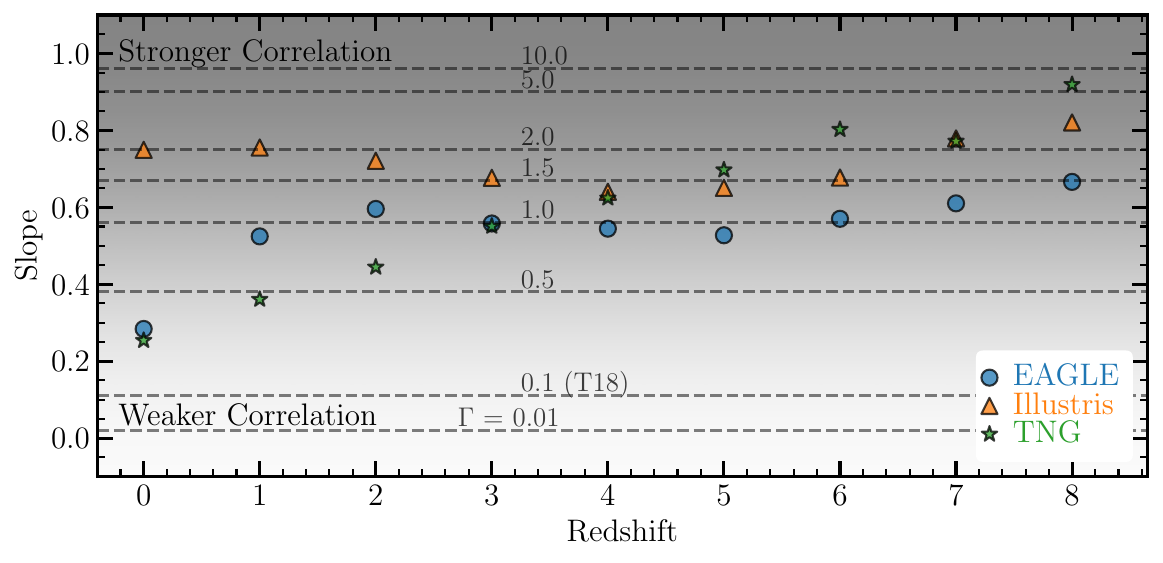}
    \caption{Slope of the offset\secondedit{s} from the MZRs for EAGLE (blue \edit{cirlces}), Illustris (orange \edit{triangles}), and TNG (green \edit{stars}) as a function of redshift with horizontal lines corresponding slopes obtained by running our analytic model (Section~\ref{subsec:analyticModel}) with certain values of $\Gamma$ (equation \ref{eqn:Gamma}). \edit{The gradient background indicates that slopes closer to unity indicate a stronger relationship between $\Delta Z_{\rm gas}$ and $\Delta Z_{*}$}. We note that the $\Gamma = 0.1$ line corresponds to predictions by \citetalias{Torrey_2018}.}
    \label{fig:slope_comparison}
\end{figure*}

% \red{Scatter -- understanding the strength of correlations also depends upon the magn of scatter. However, we find that the scatter is not important. Talk more about that. Incorporate new Figure 7, talk more about the scatter within the relation (different than the standard deviation quoted in Figure 7).}

The $M_*-Z_{\rm gas}-{\rm SFR}$ relation (i.e., FMR) follows from fairly straight-forward arguments: galactic inflows of pristine gas drive down the metallicity, but increase the available gas reservoir for star formation. 
Conversely, systems with limited or no gas inflows steadily consume their gas reservoirs while creating new metals.
Stellar metallicities, however, are not {\it directly} impacted by inflows.
Newly formed stars inherit the metallicity of the gas in which they form.
As such, the stellar metallicity represents the average gas-phase metallicity of the galaxy integrated over its entire lifetime.
It follows, then, that the gas-phase and stellar metallicity of a system \textit{may} be correlated, where the gas phase metallicity has the ability to more rapidly change in the presence of, e.g., rapid pristine gas inflows or significant enrichment events.
In this section we investigate the extent of the correlation between gas and stellar metallicities in Illustris, TNG, and EAGLE.

By defining the \MZgR{} in the same fashion as outlined in Section~\ref{subsec:stellar_MZR} for the \MZsR{} and interpolating between the mass bin centers, we obtain an offset from the median relation in both the gas-phase ($\Delta Z_{\rm gas}$) and stellar ($\Delta Z_{*}$) components for each galaxy.
We find that there is a linear relationship between these offsets in all three simulations (see Figure~\ref{fig:offset_relations} for $z=0$ offsets).
This linear relationship implies that the gas and stellar phase metallicities are tightly coupled.
We fit these offsets, $\Delta Z_{\rm gas}$ versus $\Delta Z_*$, with a linear least squares fit requiring that the intercept pass through the origin\footnote{This forced intercept is sensible since the MZRs are defined as a median, thus we expect the distributions of $\Delta Z_{*}$ and $\Delta Z_{\rm gas}$ to be centered around 0 (which, indeed, can be seen is roughly true from the histograms in the top panels of Figure~\ref{fig:offset_relations}).}.
\secondedit{As we discuss further in Section~\ref{subsec:analyticModel},} slopes approaching a value of unity (like that of Illustris' at $z=0$) indicate a more direct relationship between the offsets of the stellar and gas phase metallicities with respect to their respective \thirdedit{mass-metallicity} relationship.
%: changes in the gas phase are seen in the stars almost immediately.
Flatter slopes (e.g., EAGLE at $z=0$) indicate a weaker relationship.
%: the stars take a while to respond to changes in the gas-phase.
Similar to the slope, understanding the strength of the correlation also depends on the magnitude of scatter within the offsets.
A very tight correlation implies that stellar metallicities are \secondedit{more directly linked to} the gas-phase, while broader distributions imply that stellar metallicities \secondedit{are less directly linked to} the gas-phase.
Thus, both the slope of the offsets and scatter within the relation inform how quickly the stars can respond to perturbations within the gas-phase metallicity.
We find the slope of the offsets to be a more natural measure of the relationship here and thus use it as our primary diagnostic in this discussion.

\edit{It is important to note that while $\Delta Z_{\rm gas}$ and $\Delta Z_*$ are related to the absolute $Z_{\rm gas}$ and $Z_*$ of a system, they are different quantities.
We therefore caution the reader against taking variations in the bulk offsets from the MZRs across redshifts as representative of the difference between a single galaxy's metal evolution through two cosmic times (we discuss this subtle difference with our treatment of a toy model in Section~\ref{subsec:analyticModel}).
}

Performing the same offset analysis on at all redshifts, we obtain the time evolution shown in Figure~\ref{fig:slope_comparison}.
We find general agreement between the TNG and EAGLE simulations at $z\lesssim3$: a shallower slope that steepens with redshift.
Illustris, however, follows the opposite trend in this same redshift range, a very steep relation that weakens going back in time.
Beyond $z\gtrsim4$, all three simulations have slopes that increases with increasing redshift.

% \red{I don't think this belongs here?}
% As was shown in Section~\ref{subsec:stellar_MZR}, the three simulations' \MZsR{}'s qualitatively agree with each other in terms of general shape; 
% however, more quantitatively, the normalisations and low mass end power-law slope of the relations vary by a significant amount.
% It is therefore worth appreciating that despite the quite different normalisation, evolution, and shape of the \MZsR{} between the three simulations that the residual correlation appears in all three of them.

In spite of the $M_*-Z_*-{\rm SFR}$ having non-negligible redshift evolution in the strength of the residual correlation across redshift, the $M_*-Z_*-{\rm SFR}$ {\it is} rooted in the close relationship of gas-phase and stellar metallicities across time which is consistently present across time, albeit with some change in strength.

\section{Discussion}
\label{sec:discussion}

\subsection{Toy Model}
\label{subsec:analyticModel}

\begin{figure*}
    \centering
    \includegraphics[width=\linewidth]{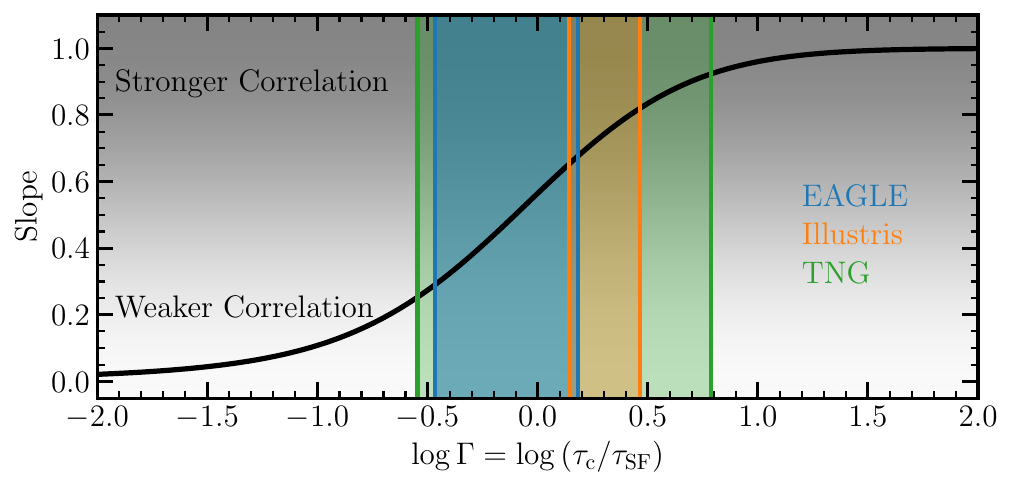}
    \caption{Slopes of the offset relation of the gas-phase and stellar metallicities (as in Figure~\ref{fig:offset_relations}) versus $\log\Gamma$, the ratio of the coherence time-scale ($\tau_{\rm c}$; the time-scale on which gas-phase metallicities are perturbed) and stellar formation time-scale ($\tau_{\rm SF}$; the inverse sSFR). 
    \edit{The black line represents predictions by our toy model (Section~\ref{subsec:analyticModel}).}
    Slopes closer to unity represent a tighter correlation between the gas-phase and stellar metallicities.
    The shaded regions correspond to the ranges of expected values of $\log\Gamma$ for EAGLE (blue), Illustris (orange), and TNG (green) based on the slopes in Figure~\ref{fig:slope_comparison}.}
    \label{fig:gamma_sSFR}
\end{figure*}

Stellar metallicities are naturally linked to gas-phase metallicities.
When stars are formed, they inherit the metallicity of the gas from which they were born. 
Thus, changes in the gas-phase metallicity will eventually impact stellar metallicities, albeit with a time delay that accounts for the fact that it takes time for a sufficient mass of new stars to be formed with the ``new'' inherited gas-phase metallicity value.
% The relationship between gas and stellar metallicity therefore depends on a few factors notably including the current stellar mass, rate of new stellar mass formation, and offset in metallicty between the gas and stellar components.
We model this behaviour with a simple analytic framework to better understand the origins of the \MZsR{} and the scatter about it.

\edit{
We start by defining $\Delta Z_*$ as the stellar metallicity of an individual galaxy, $Z_*$, subtracted from the median relation (i.e., \MZsR{}), $\langle Z_*\rangle$, at that galaxy's particular mass.
Thus, the rate of change of a galaxy's offset from the \MZsR{} can be written as
\begin{equation}
    \label{eqn:full_dZdt}
    \diff{(\Delta Z_*)}{t} = \diff{Z_*}{t} - \diff{\langle Z_*\rangle}{t}~.
\end{equation}
Equation~\ref{eqn:full_dZdt} tells us that the changes $\Delta Z_*$ are sensitive to not only changes in the absolute metallicity of a galaxy, but, critically, on the evolution of the normalisation of the \MZsR{}, as well.
From this we obtain two interesting regimes: (i) in the limit that the normalisation of the \MZsR{} does not evolve, or evolves much more slowly compared to the time-scale of stellar formation ($\tau_{\rm SF}$), $\Delta Z_*$'s are driven {\it only} by changes in the absolute metallicity and, conversely, (ii) a galaxy with no metal evolution over some time could experience a change in $\Delta Z_*$ via the overall normalisation of the \MZsR{} changing.
We begin by considering the first of the two regimes -- a steady state model in which there is no \MZsR{} evolution -- to better understand how a toy galaxy's gas-phase and stellar metallicities interact.
Then we explore the second regime and discuss the implications of the co-evolution of the \MZsR{} as a whole underneath galaxy-by-galaxy evolution. 
}

\subsubsection{Steady State Model}
\label{subsubsec:steady_state}

In principle, the rate of change of the stellar metallicity of a\edit{n individual} system should be a function of (i) the amount of stars being formed, (ii) the present gas-phase metallicity \edit{of gas forming new stars}, and (iii) the present stellar metallicity. 
To attain an analytic relation that is characterised by all three, we first define the stellar metallicity of the galaxy as the total mass of metals locked within stars, normalized by the total stellar mass:
\begin{equation}
    \label{eqn:stellarMetallicity}
    Z_* = \frac{M_{Z,*}}{M_*}~.
\end{equation}
By differentiating equation~\ref{eqn:stellarMetallicity}, and understanding that the rate of change of metals in stars is simply the gas-phase metallicity scaled by the amount (by mass) of stars that are being formed\footnote{We note that to be complete there should be some exchange of metals injected back into the ISM from the stars from supernovae and AGB winds \citep[see, e.g.,][]{Peng_Maiolino_2014}. For simplicity, and to isolate its effect, we neglect this term here. We discuss the implications of this in Section~\ref{subsubsec:limitations}.}, we obtain a straight-forward expression for the rate of change of the stellar metallicity that depends on exactly the three parameters described above,
\begin{equation}
    \label{eqn:rateOfChange}
    \diff{Z_*}{t} = \frac{1}{M_*}\diff{M_*}{t}\left(Z_{\rm gas} - Z_*\right)~.
\end{equation}
From equation~\ref{eqn:rateOfChange}, it follows that the stellar metallicity should tend towards the gas-phase metallicity value \edit{(of the star forming gas)} over time.
Moreover, the rate at which this occurs is set only by the amount of new stars being formed normalized by the total stellar mass of the system, i.e. the specific star formation rate.

\secondedit{
However, the quantity that we are comparing to in this model is the {\it offsets} in stellar metallicity, $\Delta Z_*$.
We must therefore use equation~\ref{eqn:full_dZdt} to model how the offsets in stellar metallicity respond.
Substituting the absolute metallicity with its offset and MZR terms, as well as rearranging, we are left with
\begin{equation}
\begin{split}
    \diff{(\Delta Z_*)}{t} = &\frac{1}{M_*}\diff{M_*}{t}\left[\langle Z_{\rm gas}\rangle - \langle Z_{*}\rangle\right] \\
    &+ \frac{1}{M_*}\diff{M_*}{t}\left(\Delta Z_{\rm gas} - \Delta Z_*\right) - \diff{\langle Z_*\rangle}{t}~.
    \label{eqn:fullDeltaZ}
\end{split}
\end{equation}
For now we will assume that the \MZsR{} does not evolve (${\rm d} \langle Z_* \rangle / {\rm d}t = 0$) and that the stellar and gas-phase MZRs are at the same value ($\langle Z_{\rm gas} \rangle = \langle Z_{*} \rangle$) such that the first and third term of the left-hand side are zero.
We will revisit these assumptions in Section~\ref{subsubsec:MZR_evo}.
Applying these assumptions to equation~\ref{eqn:fullDeltaZ} we are left with
\begin{equation}
    \diff{(\Delta Z_*)}{t} = \frac{1}{M_*}\diff{M_*}{t}\left(\Delta Z_{\rm gas} - \Delta Z_*\right)~.
    \label{eqn:simplifiedDeltaZ}
\end{equation}
}

With this representation of the evolution of galaxy metal content, we model the response of a toy galaxy in this paradigm.
For simplicity, we assume that our toy galaxy initially lies directly on both the \MZgR{} and \MZsR{}, so that its offsets from either are zero\footnote{\secondedit{We note that non-zero offsets in either the gas-phase or stellar metallicities (or both) do not significantly impact the results of this toy model.}}.
Next, we draw a deviation for the gas-phase metallicity \secondedit{($\Delta Z_{\rm gas}$)} from a normal distribution with a standard deviation of {0.12} dex, corresponding to a typical standard deviation of the distribution of gas-phase metallicity offsets in the simulations (see\secondedit{, e.g.,} top panels of Figure~\ref{fig:offset_relations}), centered at an offset of 0 dex.
% Physically this corresponds to gas accretion from, e.g., the CGM or an infalling galaxy.
The toy galaxy's gas-phase metallicity tends towards the randomly drawn value over a time-scale drawn from an exponential distribution with some coherence time-scale, $\tau_{\rm c}$, nominally set to one-tenth a Hubble time (as per \citetalias{Torrey_2018}; see discussion later in Section~\ref{subsubsec:limitations}).
\secondedit{Physically these perturbations are meant to mimic a potential pathway that a galaxy's gas-phase metallicity will undertake as it evolves.}
% \footnote{In our model, we ensure that the gas-phase metallicity reaches the value set by the deviation in exactly one coherence time-scale. In reality, however, it is quite possible that a system will reach the target (gas-phase) metallicity and plateau faster than one coherence time-scale, or will not reach the target metallicity at all. \red{Though, the latter is identical to drawing a lower target.}}.
As its gas-phase metallicity changes, we model the response of the toy galaxy's stellar metallicities by \secondedit{equation~\ref{eqn:simplifiedDeltaZ}}.
Specifically, we model its formation of new stars with a stellar formation time-scale, $\tau_{\rm SF}$, as the inverse of sSFRs consistent with observations (one over the hubble time).

Simply put, the coherence time-scale represents how quickly the gas-phase metallicity is perturbed, while the stellar formation time-scale informs how quickly the stars can respond to these changes.
Together the coherence and star formation time-scales fully parameterise the behavior of our toy model.
Thus, it is informative to define a (dimensionless) ratio of the two:
\begin{equation}
    \label{eqn:Gamma}
    \Gamma = \frac{\tau_{\rm c}}{\tau_{\rm SF}}~.
\end{equation}
Since, nominally, $\tau_{\rm c}$ is one-tenth a Hubble time and $\tau_{\rm SF}$ is a Hubble time, $\Gamma$ takes a fiducial value of 0.1.
It is worth noting that this $\Gamma$ holds all of the predictive power of our model.
By increasing $\Gamma$, the gas-phase metallicity perturbations take longer (or identically, the system forms stars more quickly) and thus the stellar metallicity might track the gas-phase metallicity closely.
Conversely, by decreasing $\Gamma$, we expect that the stellar metallicities take longer to equilibrate to the gas-phase metallicities, \secondedit{and therefore} the two metallicities are not likely to remain matched as the gas phase metallicity changes.
Thus, the strength of the relation of the offsets of the metallicities from their respective MZR's is a direct effect of $\Gamma$.

By scaling the initial conditions for the sSFR\footnote{Though we choose to vary sSFR, this is, in principle, exactly equivalent to decreasing the coherence time-scale. We choose the former for the sake of convenience.} we 
find the strength (i.e., slope) of the offsets from the MZR values can be uniquely described by a value of $\Gamma$ (see Figure~\ref{fig:gamma_sSFR}).
Lower (higher) $\Gamma$ corresponds to a weaker (stronger) correlation between the two metallicities.
This result agrees with na\"ive expectations: if the gas-phase metallicity evolves very quickly compared to the time it takes to form new stars, then the stellar metallicities will significantly lag behind the gas-phase metallicities.

We do note that the distribution of gas-phase metallicity offsets in the simulations more closely follows a skew-normal distribution with skewness parameters ranging from \edit{-1 to -3.5 (with the exceptions of TNG $z=2$, Illustris $z=0$, and EAGLE $z=7$ and $8$, which have skews of 0.0, -0.59, +0.36, and +0.70, respectively)}.
Despite the magnitude of skew $>1$ indicating that the distribution is non-Gaussian, when sampling gas-phase metallicity perturbations from a skew-normal distribution in our analytic model, we find that the overall slope of the correlation remains unchanged.
Further, as mentioned in Section~\ref{subsec:origin_MZR}, we forced the intercept of the offset relation to the origin; however, when sampling from a skewed normal distribution, we remove this requirement.
This is sensible since the skewed normal distribution potentially shifts the `typical' value of the distribution away from an offset of 0.
We note, however, that even if we continue to require the regression to pass through the origin, the quantitative change in the slope is modest.
Thus, since the key results found here are not significantly impacted by using a skew normal distribution, we opt to use a regular Gaussian distribution.

With the predictions offered by our analytic model and the slopes found in Section~\ref{subsec:origin_MZR}, we predict the range of values for $\Gamma$ in each simulation (Figure~\ref{fig:gamma_sSFR}).
For EAGLE, we find that the $\log\Gamma$ values range from $\sim-0.5$ to $0.2$, in TNG we find $\log\Gamma$ ranges from \edit{$\sim-0.55$ to $0.8$}, and in Illustris they range from $\sim0.2$ to $0.5$.
% We note that the range of values in Illustris is much narrower and offset from those of TNG and EAGLE.
% Since the sSFRs are comparable from simulation to simulation (i.e., Figures~\ref{fig:MZSFR*}~and~\ref{fig:MZSFR_sSFR}), this offset is necessarily
% % , based on the analytic framework outlined here,
% caused by the time-scales on which Illustris' gas-phase metals change.
% Evidently, we find that Illustris galaxies' gas-phase metallicities take longer to change (i.e., longer coherence time-scales) than their TNG and EAGLE counterparts.

\subsubsection{Impact of the underlying \texorpdfstring{MZRs}{MZRs}}
\label{subsubsec:MZR_evo}

\edit{
It is important to note that the slopes quoted in Figures~\ref{fig:offset_relations}~and~\ref{fig:slope_comparison} are based on $\Delta Z_{\rm gas}$ and $\Delta Z_{*}$ -- offsets from median relations. 
As we noted at the beginning of this discussion, changes in $\Delta Z$ are sensitive to not only changes in a system's absolute $Z$ but to changes in the underlying MZRs as well.
Yet, in the previous section, we assume that the median relation does not evolve with time \secondedit{and that the gas-phase median relation is equal to that of the stellar relation}.
We relax that assumption in this section and consider how the evolution \secondedit{and interplay} of the underlying relations impact the interpretation of our toy model.

\secondedit{
Both the first and third terms on the right-hand side of equation~\ref{eqn:fullDeltaZ} imply that a galaxy's $\Delta Z_*$ changes owing to bulk changes in the underlying median relation.
While these terms are related, each represents slightly different physical mechanisms.
The first term expresses the difference between the median \MZgR{} and median \MZsR{}.
This term is qualitatively similar to that of the offset term presented in Section~\ref{subsubsec:steady_state}: the entire \MZsR{} tends towards the \MZgR{} over some time-scale.
The larger the difference in the two median relations, the faster the \MZsR{} will respond.
Physically (both globally and individually), this links to the idea that the stellar metallicity is adopted from the gas where stars form.
The next generation of stars will have a higher metallicity when the gas-phase metals are higher than that of the stars.
On the other hand, the third term models the evolution of the normalisation of the \MZsR{}.
Individual $\Delta Z_*$s will experience changes owing to the evolution in the \MZsR{}, regardless of what the absolute stellar metallicity is doing.
Changes in the \MZsR{} are set partially by the whole relation tending towards the \MZgR{}, but also include newly formed metals within the stars of the galaxies themselves.
Importantly, both the first and third terms are sensitive to changes in the MZRs.
}

The evolution of \MZsR{} plays \secondedit{a more explicit role} in interpreting our toy model, \secondedit{we therefore isolate its effect here by only considering the second and third terms on the right-hand side of equation~\ref{eqn:fullDeltaZ}}.
The evolution of the normalisation only becomes important when it is significantly faster than stellar formation time-scales, however.
Once \MZsR{} evolution becomes significant enough, we can no longer neglect \thirdedit{these additional terms in equation~\ref{eqn:fullDeltaZ}}.
By taking the difference in metallicity within all the mass bins within the \MZsR{} at each redshift and dividing by difference in Hubble times between two given redshits, we find that the median evolution of the normalisation of the \MZsR{} is 0.08 dex/Gyr in EAGLE, 0.02 dex/Gyr in TNG, and 0.0 dex/Gyr in Illustris.
We note that these values are a crude approximation that do not account for possible mass and redshift dependencies on the evolution rate of the \MZsR{}; nevertheless, they are useful for considering the typical redshift evolution within each simulation.
Furthermore, these evolution rates reiterate the qualitative interpretations of Figure~\ref{fig:MZRallz} in Section~\ref{subsec:stellar_MZR} -- the most redshift evolution of the \MZsR{} is found within EAGLE, with modest evolution in TNG, and virtually no evolution in Illustris.

When adding the \MZsR{} evolutionary term with both the Illustris and TNG rates, we find no significant effect on the overall obtained slope of \secondedit{the} relation of $\Delta Z_{\rm gas}$ versus $\Delta Z_*$ compared \thirdedit{to} the slope from the steady state model.
For EAGLE, since the \MZsR{} normalisation changes more quickly, the evolutionary term becomes more important. 
Specifically, when using the EAGLE \MZsR{} evolution rate, $\log\Gamma\lesssim-0.25$ (corresponding to only $z=0$) is where the model begins to deviate significantly from the steady state.
In this regime, the \MZsR{} is evolving so rapidly compared to the stellar formation time-scales that the second term in Equation~\ref{eqn:full_dZdt} dominates the change in $\Delta Z_*$ over time.
The significant \MZsR{} evolution has the net effect of systematically moving the average $\Delta Z_*$ away from 0.
Yet, by definition, these $\Delta Z_*$ values should be centered around $\sim0$.
The reason behind the systematic shift is that our model only runs a single toy galaxy's metallicity track; the additional evolutionary term just represents the {\it typical} amount the \MZsR{} evolves -- the model does not actually recalculate an \MZsR{} based on a population of galaxies' stellar metallicities at each time step.
As such, fitting the relation with a line that passes through the origin (as we did in Section~\ref{subsec:origin_MZR}) produces ill-defined results.
In order to rectify the toy model in this regime, a more sophisticated approach would be required that traces evolutionary tracks of multiple systems at once all while evolving the overall normalisation of the \MZsR{} by a fixed amount at each time step, from which $\Delta Z_*$'s from the median relation could be calculated.
While it is possible to construct such a model, we favor the simplicity of the toy model detailed above in light of the relatively good job that it does reproducing the simulations' offsets in conjunction with the limited range in which such a careful tracing of \MZsR{} evolution is necessary.

\secondedit{
The first term on the right hand-side of equation~\ref{eqn:fullDeltaZ} explicitly encodes information about the difference in absolute stellar and gas-phase MZRs but also implicitly encodes information about their respective evolution.
In the limit that the neither MZR evolves, this term is just a constant that changes the rate at which $\Delta Z_*$ evolves.
A constant, non-zero offset should steepen the slope of $\Delta Z_{\rm gas}$ versus $\Delta Z_*$ for $\langle Z_{\rm gas}\rangle > \langle Z_*\rangle$ as it increases the rate of change of $\Delta Z_*$.
It should be noted that for large differences in $\langle Z_{\rm gas}\rangle$ and $\langle Z_* \rangle$ this term may play a significant role.
Both the \MZsR{} and \MZgR{} do have some evolution associated with them, however.
The median gas-phase metallicity evolutionary rates are 0.09 dex/Gyr in EAGLE, 0.08 dex/Gyr in TNG, and 0.05 dex/Gyr in Illustris (all faster than their \MZsR{} counterparts).
A complete treatment of this model would need to explicitly track the absolute metallicities of the MZRs, as well as their evolution during the toy galaxy's evolution.
Since the \MZgR{} evolutionary values are all significant, the current model presented here does not have the capabilities to perform such a task (for reasons detailed above).
}

% We start by considering the redshift evolution of the \MZgR{}.
% The normalisation of the \MZgR{} evolves with time, both in simulations \citep{Ma_2016,DeRossi_2017,Torrey_2019} and observations \citep[e.g.,][]{Savaglio_2005,Maiolino_2008,Zahid_2011,Langeroodi_2022}, specifically the normalisation decreases with increasing redshift.
\secondedit{Besides the impact on the change on $\langle Z_{\rm gas}\rangle$}, \MZgR{} evolution has no impact \secondedit{on} our toy model.
The reason that evolution of the \MZgR{} is \secondedit{otherwise} unimportant in our model is that we draw the gas-phase metallicity perturbations about the \MZgR{} itself \secondedit{($\Delta Z_{\rm gas}$ values)}, centered at an offset of 0.
Moreover, we find that the scatter about the \MZgR{} does not vary significantly across time in the simulations.
Therefore, our perturbations to the gas-phase metallicity are not sensitive to changes of the \MZgR{}.

\secondedit{Finally, we note that the underlying MZRs have some structure as a function of mass.
We make the simplifying assumption that the stellar mass of the galaxy does not evolve significantly with time.
In reality, there would likely be some contribution to $\Delta Z_{\rm *}$ owing to the galaxy increasing in stellar mass as it assembles its stellar populations.
}

In summary, the additional \MZsR{} evolutionary term may appear necessary to properly compare our toy model's predictions to the obtained slopes of the offsets from Section~\ref{subsec:origin_MZR}; however, with the exception of $z=0$ in EAGLE, there is no difference in the interpretation by including the secondary term.
\thirdedit{Regardless, we caution that the significant MZR evolution in EAGLE makes the previously derived values of $\log \Gamma$ more uncertain compared to, e.g., Illustris which has very little MZR evolution.}
\secondedit{Additionally, \thirdedit{we find that} the difference in the median MZRs increases the slope of $\Delta Z_{\rm gas}$ versus $\Delta Z_*$ when $\langle Z_{\rm gas}\rangle > \langle Z_{*}\rangle$ and is a constant offset.}
}

\subsubsection{Limitations of the model}
\label{subsubsec:limitations}

We note that there appears to be some tension with the \citetalias{Torrey_2018} model, which (effectively) predicts that $\log\Gamma=-1$ for TNG.
Recall that we neglect the return of mass and metals back into the ISM from stars dying.
\secondedit{This removes a $(1-R)$ term on the sSFR, where $R$ is the mass return fraction from stars (identically, a $1/(1-R)$ scaling on $\tau_{\rm SF}$).}
If we instead included this return fraction, assuming that is a constant over $\tau_{\rm c}$, it is equivalent to just scaling $\Gamma$ by $(1-R)$.
This does indeed shift us closer to agreement with \citetalias{Torrey_2018}, the extent of the agreement, however, depends on the adopted value of $R$.
In order to produce a slope that corresponds to a $\log\Gamma=-1$, we must assume stellar mass return values ranging from $\sim66-97$\%.
This value is much larger than current understanding predicts (i.e., $\lesssim$40\%; \citeauthor{Pillepich_2018a} \citeyear{Pillepich_2018a}; \citeauthor{Hopkins_2023} \citeyear{Hopkins_2023}).
It is also possible that \citetalias{Torrey_2018} underestimates the amount of time that a galaxy will persist above or below the MZR.
The equivalent coherence time-scale from \citetalias{Torrey_2018} (called metallicity evolution time-scale in that work) uses a Pearson correlation coefficient to describe the strength of the correlations in the offsets \secondedit{from the \MZgR{}}.
After one coherence (metallicity evolution) time-scale the strength of the correlation is weaker, but not gone entirely.
\edit{Indeed, \cite{vanLoon_2021} find that the time-scale for evolution of gas-phase metallicities is $\gg$ 1 Gyr in EAGLE, which (particularly at high redshift) is significantly greater than a tenth of the Hubble time.}
Thus, the subtle difference in interpretation of this time-scale may push the coherence time-scale (as defined in this work) to larger values, shifting the \citetalias{Torrey_2018} predictions closer towards agreement with this work.
\edit{Furthermore, the stellar formation time-scale could also be dependent on the mass of the system, as \cite{Matthee_2019} show, with less massive galaxies having more short time-scale fluctuations.}
Another possibility is that our method of randomizing metallicities isn't really a fair match to what's happening in the simulations.
For example, we model the gas-phase metallicity changing at a constant rate over the entire coherence time-scale.
The rate of change of gas-phase metallicity might change as a function of time (from, e.g., infalling material with a non-uniform density distribution).
Moreover, it is possible that after a single coherence time-scale the system's gas-phase metallicity is not {\it immediately} perturbed again, giving additional time for the stellar metallicity value to tend towards the gas-phase value.

\edit{
It is also worth noting that this model does not include every physical mechanism with which a system's metal content could evolve.
For one, the role that mergers play is underestimated by our treatment here.
While the role that infalling gas is accounted for (i.e., a possible mechanism for the gas coherence time-scale), merging galaxies also bring in stars that could change the stellar metallicity completely disjoint from new star formation. 
Dry mergers in particular would bypass both the gas-phase coherence and stellar formation time-scales entirely by changing the stellar metallicity of a system very quickly without forming new stars or perturbing the gas-phase metallicity.
\secondedit{The fraction of stars that are born ex-situ (outside of the galaxy) depends on the mass of the galaxy \citep[][]{Rodriguez-Gomez_2016,Davidson_2020}.
The ex-situ fraction is negligible for a majority of the sample analysed in this work \citep[see, e.g.,][]{Pillepich_2015} and therefore its impact on the overall stellar metallicity is likely negligible.
Past $\log M_* \gtrsim 10.2\log M_\odot$ this fraction becomes more significant, however ($>25\%$).
}
% The fraction of stars that are born ex-situ (outside of the galaxy) over a \secondedit{Milky Way mass galaxy}'s lifetime \secondedit{can be} small \citep{Pillepich_2015}, therefore the role that ex-situ stellar metallicity plays \secondedit{can sometimes be safely ignored}.
% \secondedit{However, the fraction of ex-situ stars does become larger in more massive systems \citep[up to $\sim\!50\%$ in galaxies with stellar mass 11.0 $\log M_\odot$; see][]{Rodriguez-Gomez_2016,Davidson_2020}.}
Additionally, AGN feedback plays a vital role in the evolution of metal content of high mass galaxies, yet does not fit into our toy model here.
\cite{DeRossi_2017,DeRossi_2018} and \cite{vanLoon_2021}, using the EAGLE model, show that AGN have the effect of suppressing both the gas-phase and stellar metal content in the highest mass galaxies via quenching star formation and ejecting enriched material with outflows.
These effects are thought to cause the inversion of the $M_*-Z_{\rm gas}-{\rm SFR}$ relation at the highest masses (as in \citeauthor{Yates_2012} \citeyear{Yates_2012}), and are likely contributing to the inversion of the $M_*-Z_*-{\rm SFR}$ relation in the highest mass galaxies shown in Figure~\ref{fig:MZSFR_sSFR}. 
Regardless, such quenching or outflow mechanisms are simply not taken into account in this toy model, yet it is clear that they are important to some extent in the metal evolution of the highest mass systems.
% As was shown in Figure~\ref{fig:MZSFR_sSFR}, the $Z_*-{\rm sSFR}$ relation flattens or even inverts in the highest mass galaxies.
% It is possible that AGN feedback is driving this flattening and eventual inversion of the relationship, which {\bf is not predicted by this model??}.
}

\subsection{Comparisons with other models}
\label{subsec:comparisonModels}

Just as in observations, there is limited work quantifying the $M_*-Z_*-{\rm SFR}$ relation in simulations.
%As such, there is much to be done in this space with future work in simulated galaxies.
%In this section, we briefly discuss one model that does investigate this relation and discuss possible implications in models with different feedback models.
\citeauthor{Fontanot_2021} (\citeyear{Fontanot_2021}) used the semi-analytic model GAlaxy Evolution and Assembly ({\sc gaea}) and found a residual trend between $M_*-Z_*-{\rm SFR}$.
These authors investigate the evolution of both the gas-phase and stellar metallicities through time and recover both the typical gas-phase FMR and a similar relation for stars.
Similar to the results shown in this work (see Section~\ref{subsec:residualCorrelations}; Figure~\ref{fig:alpha_redshift}), they find the $M_*-Z_*-{\rm SFR}$ correlation evolves with redshift.
Differently than this work, however, they quantify this change in the relation not in terms of a three component relation, but instead by measuring an offset from the $z=0$ \secondedit{
defined $M_*\!-Z_*\!-{\rm SFR}$ relation
\!\footnote{\secondedit{\cite{Fontanot_2021} refer to this as the ``FMR'' in their work. We refrain from using this language here to be self-consistent with previous sections of this work (see Figure~\ref{fig:alpha_redshift} and Section~\ref{subsubsec:leastScatter} for more details).}
}}.
As such, this comparison is limited to systems with overlapping masses or sSFRs at $z=0$ and the desired redshift ($z=2.24, 4.18, ~{\rm and}~ 4.88$).
In these systems, they find that \secondedit{offsets from the $z=0$ \thirdedit{$M_*\!-{Z_{*}}\!-{\rm SFR}$} relation} are a factor of $\gtrsim$2 more than in the gas-phase.
These authors consider this increased scatter evidence for significant redshift evolution of the $M_*-Z_*-{\rm SFR}$ and attribute this evolution  to the increase of baryonic mass locked in the stars as galaxies evolve. 

\edit{\cite{DeRossi_2018} also investigated the correlation of gas content within the \MZsR{} within EAGLE (though a smaller box, higher-resolution run; {\sc recal}-L025N0752, $m_{\rm baryon} = 0.226\times10^6 M_\odot$) at $z=0$.
Those authors found that the scatter of the \MZsR{} at $z=0$ is correlated with not just the sSFR, but also the gas fraction and mass-weighted age of a galaxy's stellar population, specifically that higher metallicity systems: (i) are \secondedit{older}, (ii) have a lower gas fractions, and (iii) have lower sSFRs (and vice-versa).
The inset in the right-most panel of Figure~\ref{fig:MZSFR*} shows the quantitative comparison between the high resolution EAGLE and our analysis here.
Qualitatively, these three findings corroborate what we find in Section~\ref{subsec:residualCorrelations} with the sSFR of galaxies.
While the EAGLE run we analyse has slightly elevated metalicities compared to the higher resolution run of \cite{DeRossi_2018}, the overall relationship is remarkably similar between the two runs, suggesting the $M_*-Z_*-{\rm SFR}$ relationship is not dependent on the resolution of the simulation and rather a feature of the model at least at this low redshift.
However, more robust comparisons at varying resolutions and at higher redshifts are required to confirm this.
}

\subsubsection{Implications for stellar feedback modeling}
\label{subsubec:implications}

As mentioned in Section~\ref{subsec:simulations}, all of the cosmological simulations presented in this work are of the sub-grid ISM pressurization type (i.e., \citetalias{Springel_Hernquist_2003} for Illustris and TNG, and \citetalias{Schaye_DallaVechhia_2008} for EAGLE) which are characterized by gradual (i.e. relatively non-bursty) star formation and stellar feedback.
Thus, it appears that the existence of a $M_*-Z_*-{\rm SFR}$ correlation in all of these models -- which is naturally explained by our toy model with smoothly evolving star formation rates and regularly evolving metallicities --  constitutes a reasonably generic prediction of these smooth (i.e. non-bursty) stellar feedback models.
These sub-grid implementations of the star forming ISM are not the only ways to model the ISM, however.
The Feedback in Realistic Environments (FIRE; \citeauthor{Hopkins_2014}~\citeyear{Hopkins_2014}) model, for one, attempts to more explicitly model the multi-phase ISM.
As a consequence, the star formation (and, as a result, stellar feedback) in the FIRE model can occur in large bursts -- especially in low-mass and high-redshift systems.
These large bursts of feedback can quickly remove a significant fraction of the gas from the galaxies.

The analytic model presented in Section~\ref{subsec:analyticModel} is implicitly smooth star formation: gas is accreted onto the galaxy and, over some stellar formation time-scale $\tau_{\rm SF}$, forms into stars.
While this model is certainly simplistic, it produces a sensible result in the stellar metallicity of the system lags behind the gas-phase metallicity as subsequent generations of stars form from the enriched ISM, evolve, and return the metals to form new stars.
Our model breaks down in a bursty star formation paradigm.
Rapid ejections of gas from the discs reduces the time that stars have to form as the star forming gas is removed from the system.
It is likely that these bursts would significantly disrupt the process of stellar metallicites catching up to the gas-phase values.
Thus, the strength of the residual correlation within the \MZsR{}, even if such a thing exists, is likely to be sensitive to the ``burstiness'' of stellar feedback within galaxies.
Further, these interruptions would likely leave their mark on the correlation between the gas-phase and stellar metallicities, as well, potentially weakening the overall correlation.
% as stellar metallicities would be unable to ``catch-up'' to gas-phase values.

Interestingly, the feedback prescription used in the SAM employed in \citeauthor{Fontanot_2021} (\citeyear{Fontanot_2021}; {\sc gaea}, \citeauthor{Hirschmann_2016} \citeyear{Hirschmann_2016}) takes some inspiration from the FIRE model \citep[][]{Hopkins_2014} insofar as it attempts to model bursty feedback, yet, as previously mentioned, \cite{Fontanot_2021} still recover a similar residual correlation within the \MZsR{}.
While quantitative comparisons on the strength of the correlation are difficult to make, owing to the largely different methodology in the two works, it is worth noting that the relationship indeed does exist to {\it some} extent in more bursty feedback models.
However, SAMs reconstruct feedback via approximations to observations combined with theoretical models.
Thus, it is possible that these approximations might curtail the effectiveness of the feedback compared to hydrodynamic simulations and a residual correlation may or may not appear in full hydrodynamic simulations, e.g., FIRE.

Further examination, in these hydrodynamic simulations with bursty feedback is required to fully understand the effect of burstiness on this residual trend.
However, we suggest that the (lack of) correlation between gas and stellar metallicities for galaxy populations may provide a promising opportunity for constraining or confirming the role of bursty feedback in galaxy assembly.

%\red{Maybe not strong enough here?}

%Thus, more measurements of both gas-phase and stellar metallicities for large populations of galaxies at high redshift are required to constrain current (and future) feedback models within galaxy simulations.

% If true, I think a powerful statement for observers writing proposals would be that measuring the gas and stellar metallicities for large samples at high redshift would be constraining for feedback models

\section{Conclusions}
\label{sec:conclusions}

In this work, we select star-forming, central galaxies with stellar masses $8.0 < \log(M_*~[M_\odot]) < 12.0$ and gas masses $\log(M_{\rm gas}~[M_\odot]) > 8.5$ from the Illustris, TNG, and EAGLE cosmological simulations.
Taking the global metallicity of these galaxies, we create a stellar mass-stellar metallicity relation (\MZsR{}) and stellar mass-gas-phase metallicity relation (\MZgR{}) for each at $z=0-8$.
With these, we can find each galaxy's offset from each MZR and examine residual trends.
We examine these residual trends by using a simple analytic model to track the co-evolution of the gas-phase and stellar metallicities.

Our conclusions are as follows:

\begin{itemize}
    \item The general trend of each of the \MZsR{} in each simulation is similar: less metals in low mass galaxies which increases with stellar mass before plateauing at the highest masses. The overall normalisation of the \MZsR{} (Figure~\ref{fig:MZR*}) and the redshift evolution (Figure~\ref{fig:MZRallz}), however, vary significantly in between Illustris, TNG, and EAGLE.
    \item The scatter about the \MZsR{} is correlated with the sSFR of the galaxies.
    Higher (lower) metallicity galaxies tend to have lower (higher) specific star formation rates (Figure~\ref{fig:MZSFR*}~and~\ref{fig:MZSFR_sSFR}).
    This result follows suit from observations of the scatter about the \MZgR{} (e.g., \citeauthor{Ellison_2008}~\citeyear{Ellison_2008}; \citetalias{Mannucci_2010}; \citeauthor{Lara_Lopez_2010}~\citeyear{Lara_Lopez_2010}). 
    However, we find that the dependence on SFR varies as a function of redshift (Figure~\ref{fig:alpha_redshift}) \edit{in contrast to observations and results from simulations of an ``FMR'' in the gas-phase}.
    \item The origin of both this correlated scatter and the \MZsR{} can be traced back to the intercorrelation of stellar and gas-phase metallicities, which we have fairly strong correlations within these simulations (Figures~\ref{fig:offset_relations}~and~\ref{fig:slope_comparison}).
    \item We present a simple toy model for the causal evolution of the stellar metallicities from the gas-phase metallicities.
    We determine that the driving factor between the correlation of the gas-phase and stellar metallicities depends on the interplay of the coherence time-scale (how quickly gas-phase metallicities change) and the stellar formation time-scale (how quickly new stars form).
    By definition a dimensionless parameter, $\Gamma$, as the ratio of these two time-scales, we can quantify the correlation between the gas-phase and stellar metallicities in the simulations we analyse (Figure~\ref{fig:gamma_sSFR}).
\end{itemize}

The broad-strokes agreement in the residual correlations between Illustris, TNG, and EAGLE suggest that this $M_*-Z_*-{\rm SFR}$ may be a fairly generic prediction of galaxy formation models with this ISM pressure support creating smooth star formation histories.
As the gas-phase metallicities are perturbed, the stellar metallicities have sufficient time to respond and ``catch-up'' to the gas-phase metals.
This, however, is not obviously true {\it a priori} in models with more bursty assembly histories, thus the existence of this residual correlation could potentially provide an observable diagnostic for ISM feedback models. 

\section*{Acknowledgements}

\edit{We thank the anonymous referee whose comments increased the quality of this work.
Similarly, we thank Joop Schaye for his guidance with the technical description of the EAGLE model and simulation suite.}
We acknowledge the Virgo Consortium for making their simulation data available. The EAGLE simulations were performed using the DiRAC-2 facility at Durham, managed by the ICC, and the PRACE facility Curie based in France at TGCC, CEA, Bruy\`eresle-Ch\^atel.
AMG and PT acknowledges support from NSF grant AST-2346977.
K.G. is supported by the Australian Research Council through the Discovery Early Career Researcher Award (DECRA) Fellowship (project number DE220100766) funded by the Australian Government. 
K.G. is supported by the Australian Research Council Centre of Excellence for All Sky Astrophysics in 3 Dimensions (ASTRO~3D), through project number CE170100013. 

%%%%%%%%%%%%%%%%%%%%%%%%%%%%%%%%%%%%%%%%%%%%%%%%%%
\section*{Data Availability}

All reduced data products, analysis scripts, and figures the support the conclusions of this work are all available publicly at \href{https://github.com/AlexGarcia623/interplay_gas_stars}{https://github.com/AlexGarcia623/interplay\_gas\_stars}.
Data from the Illustris and IllustrisTNG is already available on each project's respective website. 
Illustris: \href{https://www.illustris-project.org/data/}{https://www.illustris-project.org/data/} and IllustrisTNG: \href{https://www.tng-project.org/data/}{https://www.tng-project.org/data/}.
Similarly, data products from the EAGLE simulations are available for public download via the Virgo consortium’s website: \href{https://icc.dur.ac.uk/Eagle/database.php}{https://icc.dur.ac.uk/Eagle/database.php}



\bibliographystyle{mnras}
\bibliography{bibliography} % if your bibtex file is called example.bib

\begin{thebibliography}{}
\makeatletter
\relax
\def\mn@urlcharsother{\let\do\@makeother \do\$\do\&\do\#\do\^\do\_\do\%\do\~}
\def\mn@doi{\begingroup\mn@urlcharsother \@ifnextchar [ {\mn@doi@} {\mn@doi@[]}}
\def\mn@doi@[#1]#2{\def\@tempa{#1}\ifx\@tempa\@empty \href {http://dx.doi.org/#2} {doi:#2}\else \href {http://dx.doi.org/#2} {#1}\fi \endgroup}
\def\mn@eprint#1#2{\mn@eprint@#1:#2::\@nil}
\def\mn@eprint@arXiv#1{\href {http://arxiv.org/abs/#1} {{\tt arXiv:#1}}}
\def\mn@eprint@dblp#1{\href {http://dblp.uni-trier.de/rec/bibtex/#1.xml} {dblp:#1}}
\def\mn@eprint@#1:#2:#3:#4\@nil{\def\@tempa {#1}\def\@tempb {#2}\def\@tempc {#3}\ifx \@tempc \@empty \let \@tempc \@tempb \let \@tempb \@tempa \fi \ifx \@tempb \@empty \def\@tempb {arXiv}\fi \@ifundefined {mn@eprint@\@tempb}{\@tempb:\@tempc}{\expandafter \expandafter \csname mn@eprint@\@tempb\endcsname \expandafter{\@tempc}}}

\bibitem[\protect\citeauthoryear{{Andrews} \& {Martini}}{{Andrews} \& {Martini}}{2013}]{Andrews_Martini_2013}
{Andrews} B.~H.,  {Martini} P.,  2013, \mn@doi [\apj] {10.1088/0004-637X/765/2/140}, \href {https://ui.adsabs.harvard.edu/abs/2013ApJ...765..140A} {765, 140}

\bibitem[\protect\citeauthoryear{{Baker} et~al.,}{{Baker} et~al.}{2023}]{Baker_Maiolino_2023}
{Baker} W.~M.,  et~al., 2023, \mn@doi [\mnras] {10.1093/mnras/stac3594}, \href {https://ui.adsabs.harvard.edu/abs/2023MNRAS.519.1149B} {519, 1149}

\bibitem[\protect\citeauthoryear{{Beverage}, {Kriek}, {Conroy}, {Bezanson}, {Franx}  \& {van der Wel}}{{Beverage} et~al.}{2021}]{Beverage_2021}
{Beverage} A.~G.,  {Kriek} M.,  {Conroy} C.,  {Bezanson} R.,  {Franx} M.,   {van der Wel} A.,  2021, \mn@doi [\apjl] {10.3847/2041-8213/ac12cd}, \href {https://ui.adsabs.harvard.edu/abs/2021ApJ...917L...1B} {917, L1}

\bibitem[\protect\citeauthoryear{{Blanc}, {Lu}, {Benson}, {Katsianis}  \& {Barraza}}{{Blanc} et~al.}{2019}]{Blanc_2019}
{Blanc} G.~A.,  {Lu} Y.,  {Benson} A.,  {Katsianis} A.,   {Barraza} M.,  2019, \mn@doi [\apj] {10.3847/1538-4357/ab16ec}, \href {https://ui.adsabs.harvard.edu/abs/2019ApJ...877....6B} {877, 6}

\bibitem[\protect\citeauthoryear{{Bothwell}, {Maiolino}, {Kennicutt}, {Cresci}, {Mannucci}, {Marconi}  \& {Cicone}}{{Bothwell} et~al.}{2013}]{Bothwell_2013}
{Bothwell} M.~S.,  {Maiolino} R.,  {Kennicutt} R.,  {Cresci} G.,  {Mannucci} F.,  {Marconi} A.,   {Cicone} C.,  2013, \mn@doi [\mnras] {10.1093/mnras/stt817}, \href {https://ui.adsabs.harvard.edu/abs/2013MNRAS.433.1425B} {433, 1425}

\bibitem[\protect\citeauthoryear{{Bothwell}, {Maiolino}, {Peng}, {Cicone}, {Griffith}  \& {Wagg}}{{Bothwell} et~al.}{2016}]{Bothwell_2016}
{Bothwell} M.~S.,  {Maiolino} R.,  {Peng} Y.,  {Cicone} C.,  {Griffith} H.,   {Wagg} J.,  2016, \mn@doi [\mnras] {10.1093/mnras/stv2121}, \href {https://ui.adsabs.harvard.edu/abs/2016MNRAS.455.1156B} {455, 1156}

\bibitem[\protect\citeauthoryear{Chabrier}{Chabrier}{2003}]{Chabrier_2003}
Chabrier G.,  2003, \mn@doi [PASP] {10.1086/376392}, 115, 763

\bibitem[\protect\citeauthoryear{{Chisholm}, {Rigby}, {Bayliss}, {Berg}, {Dahle}, {Gladders}  \& {Sharon}}{{Chisholm} et~al.}{2019}]{Chisholm_2019}
{Chisholm} J.,  {Rigby} J.~R.,  {Bayliss} M.,  {Berg} D.~A.,  {Dahle} H.,  {Gladders} M.,   {Sharon} K.,  2019, \mn@doi [\apj] {10.3847/1538-4357/ab3104}, \href {https://ui.adsabs.harvard.edu/abs/2019ApJ...882..182C} {882, 182}

\bibitem[\protect\citeauthoryear{{Choi}, {Conroy}, {Moustakas}, {Graves}, {Holden}, {Brodwin}, {Brown}  \& {van Dokkum}}{{Choi} et~al.}{2014}]{Choi_2014}
{Choi} J.,  {Conroy} C.,  {Moustakas} J.,  {Graves} G.~J.,  {Holden} B.~P.,  {Brodwin} M.,  {Brown} M. J.~I.,   {van Dokkum} P.~G.,  2014, \mn@doi [\apj] {10.1088/0004-637X/792/2/95}, \href {https://ui.adsabs.harvard.edu/abs/2014ApJ...792...95C} {792, 95}

\bibitem[\protect\citeauthoryear{{Crain} et~al.,}{{Crain} et~al.}{2015}]{Crain_2015}
{Crain} R.~A.,  et~al., 2015, \mn@doi [\mnras] {10.1093/mnras/stv725}, \href {https://ui.adsabs.harvard.edu/abs/2015MNRAS.450.1937C} {450, 1937}

\bibitem[\protect\citeauthoryear{{Cresci}, {Mannucci}  \& {Curti}}{{Cresci} et~al.}{2019}]{Cresci_2019}
{Cresci} G.,  {Mannucci} F.,   {Curti} M.,  2019, \mn@doi [\aap] {10.1051/0004-6361/201834637}, \href {https://ui.adsabs.harvard.edu/abs/2019A&A...627A..42C} {627, A42}

\bibitem[\protect\citeauthoryear{{Cullen} et~al.,}{{Cullen} et~al.}{2019}]{Cullen_2019}
{Cullen} F.,  et~al., 2019, \mn@doi [\mnras] {10.1093/mnras/stz1402}, \href {https://ui.adsabs.harvard.edu/abs/2019MNRAS.487.2038C} {487, 2038}

\bibitem[\protect\citeauthoryear{{Cullen} et~al.,}{{Cullen} et~al.}{2021}]{Cullen_2021}
{Cullen} F.,  et~al., 2021, \mn@doi [\mnras] {10.1093/mnras/stab1340}, \href {https://ui.adsabs.harvard.edu/abs/2021MNRAS.505..903C} {505, 903}

\bibitem[\protect\citeauthoryear{{Curti}, {Mannucci}, {Cresci}  \& {Maiolino}}{{Curti} et~al.}{2020}]{Curti_2020}
{Curti} M.,  {Mannucci} F.,  {Cresci} G.,   {Maiolino} R.,  2020, \mn@doi [\mnras] {10.1093/mnras/stz2910}, \href {https://ui.adsabs.harvard.edu/abs/2020MNRAS.491..944C} {491, 944}

\bibitem[\protect\citeauthoryear{{Curti} et~al.,}{{Curti} et~al.}{2023}]{Curti_2023}
{Curti} M.,  et~al., 2023, \mn@doi [arXiv e-prints] {10.48550/arXiv.2304.08516}, \href {https://ui.adsabs.harvard.edu/abs/2023arXiv230408516C} {p. arXiv:2304.08516}

\bibitem[\protect\citeauthoryear{{D'Souza} \& {Bell}}{{D'Souza} \& {Bell}}{2018}]{DSouza_2018}
{D'Souza} R.,  {Bell} E.~F.,  2018, \mn@doi [\mnras] {10.1093/mnras/stx3081}, \href {https://ui.adsabs.harvard.edu/abs/2018MNRAS.474.5300D} {474, 5300}

\bibitem[\protect\citeauthoryear{{Dalcanton}}{{Dalcanton}}{2007}]{Dalcanton_2007}
{Dalcanton} J.~J.,  2007, \mn@doi [\apj] {10.1086/508913}, \href {https://ui.adsabs.harvard.edu/abs/2007ApJ...658..941D} {658, 941}

\bibitem[\protect\citeauthoryear{Davis, Efstathiou, Frenk  \& White}{Davis et~al.}{1985}]{Davis_1985}
Davis M.,  Efstathiou G.,  Frenk C.~S.,   White S. D.~M.,  1985, \mn@doi [ApJ] {10.1086/163168}, 292, 371

\bibitem[\protect\citeauthoryear{{Davison}, {Norris}, {Pfeffer}, {Davies}  \& {Crain}}{{Davison} et~al.}{2020}]{Davidson_2020}
{Davison} T.~A.,  {Norris} M.~A.,  {Pfeffer} J.~L.,  {Davies} J.~J.,   {Crain} R.~A.,  2020, \mn@doi [\mnras] {10.1093/mnras/staa1816}, \href {https://ui.adsabs.harvard.edu/abs/2020MNRAS.497...81D} {497, 81}

\bibitem[\protect\citeauthoryear{{De Rossi}, {Bower}, {Font}, {Schaye}  \& {Theuns}}{{De Rossi} et~al.}{2017}]{DeRossi_2017}
{De Rossi} M.~E.,  {Bower} R.~G.,  {Font} A.~S.,  {Schaye} J.,   {Theuns} T.,  2017, \mn@doi [\mnras] {10.1093/mnras/stx2158}, \href {https://ui.adsabs.harvard.edu/abs/2017MNRAS.472.3354D} {472, 3354}

\bibitem[\protect\citeauthoryear{{De Rossi}, {Bower}, {Font}  \& {Schaye}}{{De Rossi} et~al.}{2018}]{DeRossi_2018}
{De Rossi} M.~E.,  {Bower} R.~G.,  {Font} A.~S.,   {Schaye} T.,  2018, \mn@doi [Boletin de la Asociacion Argentina de Astronomia La Plata Argentina] {10.48550/arXiv.1805.06119}, \href {https://ui.adsabs.harvard.edu/abs/2018BAAA...60..121R} {60, 121}

\bibitem[\protect\citeauthoryear{Dolag, Borgani, Murante  \& Springel}{Dolag et~al.}{2009}]{Dolag_2009}
Dolag K.,  Borgani S.,  Murante G.,   Springel V.,  2009, \mn@doi [MNRAS] {10.1111/j.1365-2966.2009.15034.x}, 399, 497

\bibitem[\protect\citeauthoryear{Donnari et~al.,}{Donnari et~al.}{2019}]{Donnari_2019}
Donnari M.,  et~al., 2019, \mn@doi [MNRAS] {10.1093/mnras/stz712}, 485, 4817

\bibitem[\protect\citeauthoryear{{Ellison}, {Patton}, {Simard}  \& {McConnachie}}{{Ellison} et~al.}{2008}]{Ellison_2008}
{Ellison} S.~L.,  {Patton} D.~R.,  {Simard} L.,   {McConnachie} A.~W.,  2008, \mn@doi [\apjl] {10.1086/527296}, \href {https://ui.adsabs.harvard.edu/abs/2008ApJ...672L.107E} {672, L107}

\bibitem[\protect\citeauthoryear{Elmegreen}{Elmegreen}{1999}]{Elmegreen_1999}
Elmegreen B.~G.,  1999, \mn@doi [ApJ] {10.1086/308073}, 527, 266

\bibitem[\protect\citeauthoryear{{Faucher-Gigu{\`e}re}}{{Faucher-Gigu{\`e}re}}{2018}]{Faucher-Giguere_2018}
{Faucher-Gigu{\`e}re} C.-A.,  2018, \mn@doi [\mnras] {10.1093/mnras/stx2595}, \href {https://ui.adsabs.harvard.edu/abs/2018MNRAS.473.3717F} {473, 3717}

\bibitem[\protect\citeauthoryear{{Fontanot} et~al.,}{{Fontanot} et~al.}{2021}]{Fontanot_2021}
{Fontanot} F.,  et~al., 2021, \mn@doi [\mnras] {10.1093/mnras/stab1213}, \href {https://ui.adsabs.harvard.edu/abs/2021MNRAS.504.4481F} {504, 4481}

\bibitem[\protect\citeauthoryear{{Fraser-McKelvie} et~al.,}{{Fraser-McKelvie} et~al.}{2022}]{Fraser-McKelvia_2022}
{Fraser-McKelvie} A.,  et~al., 2022, \mn@doi [\mnras] {10.1093/mnras/stab3430}, \href {https://ui.adsabs.harvard.edu/abs/2022MNRAS.510..320F} {510, 320}

\bibitem[\protect\citeauthoryear{{Gallazzi}, {Charlot}, {Brinchmann}, {White}  \& {Tremonti}}{{Gallazzi} et~al.}{2005}]{Gallazzi_2005}
{Gallazzi} A.,  {Charlot} S.,  {Brinchmann} J.,  {White} S. D.~M.,   {Tremonti} C.~A.,  2005, \mn@doi [\mnras] {10.1111/j.1365-2966.2005.09321.x}, \href {https://ui.adsabs.harvard.edu/abs/2005MNRAS.362...41G} {362, 41}

\bibitem[\protect\citeauthoryear{{Gallazzi}, {Bell}, {Zibetti}, {Brinchmann}  \& {Kelson}}{{Gallazzi} et~al.}{2014}]{Gallazzi_2014}
{Gallazzi} A.,  {Bell} E.~F.,  {Zibetti} S.,  {Brinchmann} J.,   {Kelson} D.~D.,  2014, \mn@doi [\apj] {10.1088/0004-637X/788/1/72}, \href {https://ui.adsabs.harvard.edu/abs/2014ApJ...788...72G} {788, 72}

\bibitem[\protect\citeauthoryear{{Garcia} et~al.,}{{Garcia} et~al.}{2023}]{Garcia_2023}
{Garcia} A.~M.,  et~al., 2023, \mn@doi [\mnras] {10.1093/mnras/stac3749}, \href {https://ui.adsabs.harvard.edu/abs/2023MNRAS.519.4716G} {519, 4716}

\bibitem[\protect\citeauthoryear{{Genel} et~al.,}{{Genel} et~al.}{2014}]{Genel_2014}
{Genel} S.,  et~al., 2014, \mn@doi [\mnras] {10.1093/mnras/stu1654}, \href {https://ui.adsabs.harvard.edu/abs/2014MNRAS.445..175G} {445, 175}

\bibitem[\protect\citeauthoryear{{Gonz{\'a}lez Delgado} et~al.,}{{Gonz{\'a}lez Delgado} et~al.}{2014}]{Gonzalez_Delgado_2014}
{Gonz{\'a}lez Delgado} R.~M.,  et~al., 2014, \mn@doi [\apjl] {10.1088/2041-8205/791/1/L16}, \href {https://ui.adsabs.harvard.edu/abs/2014ApJ...791L..16G} {791, L16}

\bibitem[\protect\citeauthoryear{{Greener} et~al.,}{{Greener} et~al.}{2022}]{Greener_2022}
{Greener} M.~J.,  et~al., 2022, \mn@doi [\mnras] {10.1093/mnras/stac2355}, \href {https://ui.adsabs.harvard.edu/abs/2022MNRAS.516.1275G} {516, 1275}

\bibitem[\protect\citeauthoryear{{Guidi}, {Scannapieco}, {Walcher}  \& {Gallazzi}}{{Guidi} et~al.}{2016}]{Guidi_2016}
{Guidi} G.,  {Scannapieco} C.,  {Walcher} J.,   {Gallazzi} A.,  2016, \mn@doi [\mnras] {10.1093/mnras/stw1790}, \href {https://ui.adsabs.harvard.edu/abs/2016MNRAS.462.2046G} {462, 2046}

\bibitem[\protect\citeauthoryear{{Halliday} et~al.,}{{Halliday} et~al.}{2008}]{Halliday_2008}
{Halliday} C.,  et~al., 2008, \mn@doi [\aap] {10.1051/0004-6361:20078673}, \href {https://ui.adsabs.harvard.edu/abs/2008A&A...479..417H} {479, 417}

\bibitem[\protect\citeauthoryear{{Hemler} et~al.,}{{Hemler} et~al.}{2021}]{Hemler_2021}
{Hemler} Z.~S.,  et~al., 2021, \mn@doi [\mnras] {10.1093/mnras/stab1803}, \href {https://ui.adsabs.harvard.edu/abs/2021MNRAS.506.3024H} {506, 3024}

\bibitem[\protect\citeauthoryear{{Hirschmann}, {De Lucia}  \& {Fontanot}}{{Hirschmann} et~al.}{2016}]{Hirschmann_2016}
{Hirschmann} M.,  {De Lucia} G.,   {Fontanot} F.,  2016, \mn@doi [\mnras] {10.1093/mnras/stw1318}, \href {https://ui.adsabs.harvard.edu/abs/2016MNRAS.461.1760H} {461, 1760}

\bibitem[\protect\citeauthoryear{{Hopkins}, {Kere{\v{s}}}, {O{\~n}orbe}, {Faucher-Gigu{\`e}re}, {Quataert}, {Murray}  \& {Bullock}}{{Hopkins} et~al.}{2014}]{Hopkins_2014}
{Hopkins} P.~F.,  {Kere{\v{s}}} D.,  {O{\~n}orbe} J.,  {Faucher-Gigu{\`e}re} C.-A.,  {Quataert} E.,  {Murray} N.,   {Bullock} J.~S.,  2014, \mn@doi [\mnras] {10.1093/mnras/stu1738}, \href {https://ui.adsabs.harvard.edu/abs/2014MNRAS.445..581H} {445, 581}

\bibitem[\protect\citeauthoryear{{Hopkins} et~al.,}{{Hopkins} et~al.}{2023}]{Hopkins_2023}
{Hopkins} P.~F.,  et~al., 2023, \mn@doi [\mnras] {10.1093/mnras/stac3489}, \href {https://ui.adsabs.harvard.edu/abs/2023MNRAS.519.3154H} {519, 3154}

\bibitem[\protect\citeauthoryear{{Huang} et~al.,}{{Huang} et~al.}{2019}]{Huang_2019}
{Huang} C.,  et~al., 2019, \mn@doi [\apj] {10.3847/1538-4357/ab4902}, \href {https://ui.adsabs.harvard.edu/abs/2019ApJ...886...31H} {886, 31}

\bibitem[\protect\citeauthoryear{{Kashino} et~al.,}{{Kashino} et~al.}{2022}]{Kashino_2022}
{Kashino} D.,  et~al., 2022, \mn@doi [\apj] {10.3847/1538-4357/ac399e}, \href {https://ui.adsabs.harvard.edu/abs/2022ApJ...925...82K} {925, 82}

\bibitem[\protect\citeauthoryear{{Kennicutt}}{{Kennicutt}}{1998}]{Kennicutt_1998}
{Kennicutt} Robert~C. J.,  1998, \mn@doi [\apj] {10.1086/305588}, \href {https://ui.adsabs.harvard.edu/abs/1998ApJ...498..541K} {498, 541}

\bibitem[\protect\citeauthoryear{{Kere{\v{s}}}, {Katz}, {Weinberg}  \& {Dav{\'e}}}{{Kere{\v{s}}} et~al.}{2005}]{Keres_2005}
{Kere{\v{s}}} D.,  {Katz} N.,  {Weinberg} D.~H.,   {Dav{\'e}} R.,  2005, \mn@doi [\mnras] {10.1111/j.1365-2966.2005.09451.x}, \href {https://ui.adsabs.harvard.edu/abs/2005MNRAS.363....2K} {363, 2}

\bibitem[\protect\citeauthoryear{{Kere{\v{s}}}, {Vogelsberger}, {Sijacki}, {Springel}  \& {Hernquist}}{{Kere{\v{s}}} et~al.}{2012}]{Keres_2012}
{Kere{\v{s}}} D.,  {Vogelsberger} M.,  {Sijacki} D.,  {Springel} V.,   {Hernquist} L.,  2012, \mn@doi [\mnras] {10.1111/j.1365-2966.2012.21548.x}, \href {https://ui.adsabs.harvard.edu/abs/2012MNRAS.425.2027K} {425, 2027}

\bibitem[\protect\citeauthoryear{{Kewley} \& {Ellison}}{{Kewley} \& {Ellison}}{2008}]{Kewley_Ellison_2008}
{Kewley} L.~J.,  {Ellison} S.~L.,  2008, \mn@doi [\apj] {10.1086/587500}, \href {https://ui.adsabs.harvard.edu/abs/2008ApJ...681.1183K} {681, 1183}

\bibitem[\protect\citeauthoryear{{Kewley}, {Nicholls}  \& {Sutherland}}{{Kewley} et~al.}{2019}]{Kewley_2019}
{Kewley} L.~J.,  {Nicholls} D.~C.,   {Sutherland} R.~S.,  2019, \mn@doi [\araa] {10.1146/annurev-astro-081817-051832}, \href {https://ui.adsabs.harvard.edu/abs/2019ARA&A..57..511K} {57, 511}

\bibitem[\protect\citeauthoryear{{Kobayashi}, {Karakas}  \& {Lugaro}}{{Kobayashi} et~al.}{2020}]{Kobayashi_2020}
{Kobayashi} C.,  {Karakas} A.~I.,   {Lugaro} M.,  2020, \mn@doi [\apj] {10.3847/1538-4357/abae65}, \href {https://ui.adsabs.harvard.edu/abs/2020ApJ...900..179K} {900, 179}

\bibitem[\protect\citeauthoryear{{Kriek} et~al.,}{{Kriek} et~al.}{2019}]{Kriek_2019}
{Kriek} M.,  et~al., 2019, \mn@doi [\apjl] {10.3847/2041-8213/ab2e75}, \href {https://ui.adsabs.harvard.edu/abs/2019ApJ...880L..31K} {880, L31}

\bibitem[\protect\citeauthoryear{{Langeroodi} et~al.,}{{Langeroodi} et~al.}{2022}]{Langeroodi_2022}
{Langeroodi} D.,  et~al., 2022, \mn@doi [arXiv e-prints] {10.48550/arXiv.2212.02491}, \href {https://ui.adsabs.harvard.edu/abs/2022arXiv221202491L} {p. arXiv:2212.02491}

\bibitem[\protect\citeauthoryear{{Lara-L{\'o}pez} et~al.,}{{Lara-L{\'o}pez} et~al.}{2010}]{Lara_Lopez_2010}
{Lara-L{\'o}pez} M.~A.,  et~al., 2010, \mn@doi [\aap] {10.1051/0004-6361/201014803}, \href {https://ui.adsabs.harvard.edu/abs/2010A&A...521L..53L} {521, L53}

\bibitem[\protect\citeauthoryear{{Lara-L{\'o}pez}, {L{\'o}pez-S{\'a}nchez}  \& {Hopkins}}{{Lara-L{\'o}pez} et~al.}{2013}]{Lara_Lopez_2013}
{Lara-L{\'o}pez} M.~A.,  {L{\'o}pez-S{\'a}nchez} {\'A}.~R.,   {Hopkins} A.~M.,  2013, \mn@doi [\apj] {10.1088/0004-637X/764/2/178}, \href {https://ui.adsabs.harvard.edu/abs/2013ApJ...764..178L} {764, 178}

\bibitem[\protect\citeauthoryear{{Lazar} et~al.,}{{Lazar} et~al.}{2020}]{Lazar_2020}
{Lazar} A.,  et~al., 2020, \mn@doi [\mnras] {10.1093/mnras/staa2101}, \href {https://ui.adsabs.harvard.edu/abs/2020MNRAS.497.2393L} {497, 2393}

\bibitem[\protect\citeauthoryear{{Lee}, {Skillman}, {Cannon}, {Jackson}, {Gehrz}, {Polomski}  \& {Woodward}}{{Lee} et~al.}{2006}]{Lee_2006}
{Lee} H.,  {Skillman} E.~D.,  {Cannon} J.~M.,  {Jackson} D.~C.,  {Gehrz} R.~D.,  {Polomski} E.~F.,   {Woodward} C.~E.,  2006, \mn@doi [\apj] {10.1086/505573}, \href {https://ui.adsabs.harvard.edu/abs/2006ApJ...647..970L} {647, 970}

\bibitem[\protect\citeauthoryear{{Leethochawalit}, {Kirby}, {Moran}, {Ellis}  \& {Treu}}{{Leethochawalit} et~al.}{2018}]{Leethochawalit_2018}
{Leethochawalit} N.,  {Kirby} E.~N.,  {Moran} S.~M.,  {Ellis} R.~S.,   {Treu} T.,  2018, \mn@doi [\apj] {10.3847/1538-4357/aab26a}, \href {https://ui.adsabs.harvard.edu/abs/2018ApJ...856...15L} {856, 15}

\bibitem[\protect\citeauthoryear{{Lian}, {Thomas}, {Maraston}, {Goddard}, {Comparat}, {Gonzalez-Perez}  \& {Ventura}}{{Lian} et~al.}{2018}]{Lian_2018}
{Lian} J.,  {Thomas} D.,  {Maraston} C.,  {Goddard} D.,  {Comparat} J.,  {Gonzalez-Perez} V.,   {Ventura} P.,  2018, \mn@doi [\mnras] {10.1093/mnras/stx2829}, \href {https://ui.adsabs.harvard.edu/abs/2018MNRAS.474.1143L} {474, 1143}

\bibitem[\protect\citeauthoryear{{Ma}, {Hopkins}, {Faucher-Gigu{\`e}re}, {Zolman}, {Muratov}, {Kere{\v{s}}}  \& {Quataert}}{{Ma} et~al.}{2016}]{Ma_2016}
{Ma} X.,  {Hopkins} P.~F.,  {Faucher-Gigu{\`e}re} C.-A.,  {Zolman} N.,  {Muratov} A.~L.,  {Kere{\v{s}}} D.,   {Quataert} E.,  2016, \mn@doi [\mnras] {10.1093/mnras/stv2659}, \href {https://ui.adsabs.harvard.edu/abs/2016MNRAS.456.2140M} {456, 2140}

\bibitem[\protect\citeauthoryear{{Maiolino} et~al.,}{{Maiolino} et~al.}{2008}]{Maiolino_2008}
{Maiolino} R.,  et~al., 2008, \mn@doi [\aap] {10.1051/0004-6361:200809678}, \href {https://ui.adsabs.harvard.edu/abs/2008A&A...488..463M} {488, 463}

\bibitem[\protect\citeauthoryear{{Mannucci}, {Cresci}, {Maiolino}, {Marconi}  \& {Gnerucci}}{{Mannucci} et~al.}{2010}]{Mannucci_2010}
{Mannucci} F.,  {Cresci} G.,  {Maiolino} R.,  {Marconi} A.,   {Gnerucci} A.,  2010, \mn@doi [\mnras] {10.1111/j.1365-2966.2010.17291.x}, \href {https://ui.adsabs.harvard.edu/abs/2010MNRAS.408.2115M} {408, 2115}

\bibitem[\protect\citeauthoryear{{Marinacci} et~al.,}{{Marinacci} et~al.}{2018}]{Marinacci_2018}
{Marinacci} F.,  et~al., 2018, \mn@doi [\mnras] {10.1093/mnras/sty2206}, \href {https://ui.adsabs.harvard.edu/abs/2018MNRAS.480.5113M} {480, 5113}

\bibitem[\protect\citeauthoryear{{Matthee} \& {Schaye}}{{Matthee} \& {Schaye}}{2019}]{Matthee_2019}
{Matthee} J.,  {Schaye} J.,  2019, \mn@doi [\mnras] {10.1093/mnras/stz030}, \href {https://ui.adsabs.harvard.edu/abs/2019MNRAS.484..915M} {484, 915}

\bibitem[\protect\citeauthoryear{{McAlpine} et~al.,}{{McAlpine} et~al.}{2016}]{McAlpine_2016}
{McAlpine} S.,  et~al., 2016, \mn@doi [Astronomy and Computing] {10.1016/j.ascom.2016.02.004}, \href {https://ui.adsabs.harvard.edu/abs/2016A&C....15...72M} {15, 72}

\bibitem[\protect\citeauthoryear{{Naiman} et~al.,}{{Naiman} et~al.}{2018}]{Naiman_2018}
{Naiman} J.~P.,  et~al., 2018, \mn@doi [\mnras] {10.1093/mnras/sty618}, \href {https://ui.adsabs.harvard.edu/abs/2018MNRAS.477.1206N} {477, 1206}

\bibitem[\protect\citeauthoryear{{Nelson} et~al.,}{{Nelson} et~al.}{2018}]{Nelson_2018}
{Nelson} D.,  et~al., 2018, \mn@doi [\mnras] {10.1093/mnras/stx3040}, \href {https://ui.adsabs.harvard.edu/abs/2018MNRAS.475..624N} {475, 624}

\bibitem[\protect\citeauthoryear{{Nelson} et~al.,}{{Nelson} et~al.}{2019a}]{Nelson_2019a}
{Nelson} D.,  et~al., 2019a, \mn@doi [Computational Astrophysics and Cosmology] {10.1186/s40668-019-0028-x}, \href {https://ui.adsabs.harvard.edu/abs/2019ComAC...6....2N} {6, 2}

\bibitem[\protect\citeauthoryear{{Nelson} et~al.,}{{Nelson} et~al.}{2019b}]{Nelson_2019b}
{Nelson} D.,  et~al., 2019b, \mn@doi [\mnras] {10.1093/mnras/stz2306}, \href {https://ui.adsabs.harvard.edu/abs/2019MNRAS.490.3234N} {490, 3234}

\bibitem[\protect\citeauthoryear{Nelson et~al.,}{Nelson et~al.}{2021}]{Nelson_2021}
Nelson E.~J.,  et~al., 2021, \mn@doi [MNRAS] {10.1093/mnras/stab2131}, 508, 219

\bibitem[\protect\citeauthoryear{{Panter}, {Jimenez}, {Heavens}  \& {Charlot}}{{Panter} et~al.}{2008}]{Panter_2008}
{Panter} B.,  {Jimenez} R.,  {Heavens} A.~F.,   {Charlot} S.,  2008, \mn@doi [\mnras] {10.1111/j.1365-2966.2008.13981.x}, \href {https://ui.adsabs.harvard.edu/abs/2008MNRAS.391.1117P} {391, 1117}

\bibitem[\protect\citeauthoryear{{Pasquali}, {Gallazzi}  \& {van den Bosch}}{{Pasquali} et~al.}{2012}]{Pasquali_2012}
{Pasquali} A.,  {Gallazzi} A.,   {van den Bosch} F.~C.,  2012, \mn@doi [\mnras] {10.1111/j.1365-2966.2012.21454.x}, \href {https://ui.adsabs.harvard.edu/abs/2012MNRAS.425..273P} {425, 273}

\bibitem[\protect\citeauthoryear{{Patr{\'\i}cio} et~al.,}{{Patr{\'\i}cio} et~al.}{2016}]{Patricio_2016}
{Patr{\'\i}cio} V.,  et~al., 2016, \mn@doi [\mnras] {10.1093/mnras/stv2859}, \href {https://ui.adsabs.harvard.edu/abs/2016MNRAS.456.4191P} {456, 4191}

\bibitem[\protect\citeauthoryear{{Peng} \& {Maiolino}}{{Peng} \& {Maiolino}}{2014}]{Peng_Maiolino_2014}
{Peng} Y.-j.,  {Maiolino} R.,  2014, \mn@doi [\mnras] {10.1093/mnras/stu1288}, \href {https://ui.adsabs.harvard.edu/abs/2014MNRAS.443.3643P} {443, 3643}

\bibitem[\protect\citeauthoryear{{Pillepich}, {Madau}  \& {Mayer}}{{Pillepich} et~al.}{2015}]{Pillepich_2015}
{Pillepich} A.,  {Madau} P.,   {Mayer} L.,  2015, \mn@doi [\apj] {10.1088/0004-637X/799/2/184}, \href {https://ui.adsabs.harvard.edu/abs/2015ApJ...799..184P} {799, 184}

\bibitem[\protect\citeauthoryear{Pillepich et~al.,}{Pillepich et~al.}{2018a}]{Pillepich_2018a}
Pillepich A.,  et~al., 2018a, \mn@doi [MNRAS] {10.1093/mnras/stx2656}, 473, 4077

\bibitem[\protect\citeauthoryear{{Pillepich} et~al.,}{{Pillepich} et~al.}{2018b}]{Pillepich_2018b}
{Pillepich} A.,  et~al., 2018b, \mn@doi [\mnras] {10.1093/mnras/stx3112}, \href {https://ui.adsabs.harvard.edu/abs/2018MNRAS.475..648P} {475, 648}

\bibitem[\protect\citeauthoryear{Pillepich et~al.,}{Pillepich et~al.}{2019}]{Pillepich_2019}
Pillepich A.,  et~al., 2019, \mn@doi [MNRAS] {10.1093/mnras/stz2338}, 490, 3196

\bibitem[\protect\citeauthoryear{{Rodriguez-Gomez} et~al.,}{{Rodriguez-Gomez} et~al.}{2016}]{Rodriguez-Gomez_2016}
{Rodriguez-Gomez} V.,  et~al., 2016, \mn@doi [\mnras] {10.1093/mnras/stw456}, \href {https://ui.adsabs.harvard.edu/abs/2016MNRAS.458.2371R} {458, 2371}

\bibitem[\protect\citeauthoryear{{Sanders}, {Shapley}, {Topping}, {Reddy}  \& {Brammer}}{{Sanders} et~al.}{2023}]{Sanders_2023}
{Sanders} R.~L.,  {Shapley} A.~E.,  {Topping} M.~W.,  {Reddy} N.~A.,   {Brammer} G.~B.,  2023, \mn@doi [arXiv e-prints] {10.48550/arXiv.2303.08149}, \href {https://ui.adsabs.harvard.edu/abs/2023arXiv230308149S} {p. arXiv:2303.08149}

\bibitem[\protect\citeauthoryear{{Savaglio} et~al.,}{{Savaglio} et~al.}{2005}]{Savaglio_2005}
{Savaglio} S.,  et~al., 2005, \mn@doi [\apj] {10.1086/497331}, \href {https://ui.adsabs.harvard.edu/abs/2005ApJ...635..260S} {635, 260}

\bibitem[\protect\citeauthoryear{{Schaller}, {Dalla Vecchia}, {Schaye}, {Bower}, {Theuns}, {Crain}, {Furlong}  \& {McCarthy}}{{Schaller} et~al.}{2015}]{Schaller_2015}
{Schaller} M.,  {Dalla Vecchia} C.,  {Schaye} J.,  {Bower} R.~G.,  {Theuns} T.,  {Crain} R.~A.,  {Furlong} M.,   {McCarthy} I.~G.,  2015, \mn@doi [\mnras] {10.1093/mnras/stv2169}, \href {https://ui.adsabs.harvard.edu/abs/2015MNRAS.454.2277S} {454, 2277}

\bibitem[\protect\citeauthoryear{{Schaye}}{{Schaye}}{2004}]{Schaye_2004}
{Schaye} J.,  2004, \mn@doi [\apj] {10.1086/421232}, \href {https://ui.adsabs.harvard.edu/abs/2004ApJ...609..667S} {609, 667}

\bibitem[\protect\citeauthoryear{Schaye \& Dalla~Vecchia}{Schaye \& Dalla~Vecchia}{2008}]{Schaye_DallaVechhia_2008}
Schaye J.,  Dalla~Vecchia C.,  2008, \mn@doi [MNRAS] {10.1111/j.1365-2966.2007.12639.x}, 383, 1210

\bibitem[\protect\citeauthoryear{{Schaye} et~al.,}{{Schaye} et~al.}{2015}]{Schaye_2015}
{Schaye} J.,  et~al., 2015, \mn@doi [\mnras] {10.1093/mnras/stu2058}, \href {https://ui.adsabs.harvard.edu/abs/2015MNRAS.446..521S} {446, 521}

\bibitem[\protect\citeauthoryear{{Schmidt}}{{Schmidt}}{1959}]{Schmidt_1959}
{Schmidt} M.,  1959, \mn@doi [\apj] {10.1086/146614}, \href {https://ui.adsabs.harvard.edu/abs/1959ApJ...129..243S} {129, 243}

\bibitem[\protect\citeauthoryear{{Sextl}, {Kudritzki}, {Zahid}  \& {Ho}}{{Sextl} et~al.}{2023}]{Sextl_2023}
{Sextl} E.,  {Kudritzki} R.-P.,  {Zahid} H.~J.,   {Ho} I.-T.,  2023, \mn@doi [arXiv e-prints] {10.48550/arXiv.2303.11024}, \href {https://ui.adsabs.harvard.edu/abs/2023arXiv230311024S} {p. arXiv:2303.11024}

\bibitem[\protect\citeauthoryear{{Shapley}, {Reddy}, {Sanders}, {Topping}  \& {Brammer}}{{Shapley} et~al.}{2023}]{Shapley_2023}
{Shapley} A.~E.,  {Reddy} N.~A.,  {Sanders} R.~L.,  {Topping} M.~W.,   {Brammer} G.~B.,  2023, \mn@doi [arXiv e-prints] {10.48550/arXiv.2303.00410}, \href {https://ui.adsabs.harvard.edu/abs/2023arXiv230300410S} {p. arXiv:2303.00410}

\bibitem[\protect\citeauthoryear{{Sijacki}, {Vogelsberger}, {Kere{\v{s}}}, {Springel}  \& {Hernquist}}{{Sijacki} et~al.}{2012}]{Sijacki_2012}
{Sijacki} D.,  {Vogelsberger} M.,  {Kere{\v{s}}} D.,  {Springel} V.,   {Hernquist} L.,  2012, \mn@doi [\mnras] {10.1111/j.1365-2966.2012.21466.x}, \href {https://ui.adsabs.harvard.edu/abs/2012MNRAS.424.2999S} {424, 2999}

\bibitem[\protect\citeauthoryear{{Sommariva}, {Mannucci}, {Cresci}, {Maiolino}, {Marconi}, {Nagao}, {Baroni}  \& {Grazian}}{{Sommariva} et~al.}{2012}]{Sommariva_2012}
{Sommariva} V.,  {Mannucci} F.,  {Cresci} G.,  {Maiolino} R.,  {Marconi} A.,  {Nagao} T.,  {Baroni} A.,   {Grazian} A.,  2012, \mn@doi [\aap] {10.1051/0004-6361/201118134}, \href {https://ui.adsabs.harvard.edu/abs/2012A&A...539A.136S} {539, A136}

\bibitem[\protect\citeauthoryear{{Springel}}{{Springel}}{2005}]{Springel_2005}
{Springel} V.,  2005, \mn@doi [\mnras] {10.1111/j.1365-2966.2005.09655.x}, \href {https://ui.adsabs.harvard.edu/abs/2005MNRAS.364.1105S} {364, 1105}

\bibitem[\protect\citeauthoryear{{Springel}}{{Springel}}{2010}]{Springel_2010}
{Springel} V.,  2010, \mn@doi [\mnras] {10.1111/j.1365-2966.2009.15715.x}, \href {https://ui.adsabs.harvard.edu/abs/2010MNRAS.401..791S} {401, 791}

\bibitem[\protect\citeauthoryear{{Springel} \& {Hernquist}}{{Springel} \& {Hernquist}}{2003}]{Springel_Hernquist_2003}
{Springel} V.,  {Hernquist} L.,  2003, \mn@doi [\mnras] {10.1046/j.1365-8711.2003.06206.x}, \href {https://ui.adsabs.harvard.edu/abs/2003MNRAS.339..289S} {339, 289}

\bibitem[\protect\citeauthoryear{Springel, White  \& Hernquist}{Springel et~al.}{2001}]{Springel_2001}
Springel V.,  White M.,   Hernquist L.,  2001, \mn@doi [ApJ] {10.1086/319473}, 549, 681

\bibitem[\protect\citeauthoryear{{Springel} et~al.,}{{Springel} et~al.}{2018}]{Springel_2018}
{Springel} V.,  et~al., 2018, \mn@doi [\mnras] {10.1093/mnras/stx3304}, \href {https://ui.adsabs.harvard.edu/abs/2018MNRAS.475..676S} {475, 676}

\bibitem[\protect\citeauthoryear{{Topping}, {Shapley}, {Reddy}, {Sanders}, {Coil}, {Kriek}, {Mobasher}  \& {Siana}}{{Topping} et~al.}{2020}]{Topping_2020}
{Topping} M.~W.,  {Shapley} A.~E.,  {Reddy} N.~A.,  {Sanders} R.~L.,  {Coil} A.~L.,  {Kriek} M.,  {Mobasher} B.,   {Siana} B.,  2020, \mn@doi [\mnras] {10.1093/mnras/staa2941}, \href {https://ui.adsabs.harvard.edu/abs/2020MNRAS.499.1652T} {499, 1652}

\bibitem[\protect\citeauthoryear{{Torrey}, {Vogelsberger}, {Sijacki}, {Springel}  \& {Hernquist}}{{Torrey} et~al.}{2012}]{Torrey_2012}
{Torrey} P.,  {Vogelsberger} M.,  {Sijacki} D.,  {Springel} V.,   {Hernquist} L.,  2012, \mn@doi [\mnras] {10.1111/j.1365-2966.2012.22082.x}, \href {https://ui.adsabs.harvard.edu/abs/2012MNRAS.427.2224T} {427, 2224}

\bibitem[\protect\citeauthoryear{{Torrey}, {Vogelsberger}, {Genel}, {Sijacki}, {Springel}  \& {Hernquist}}{{Torrey} et~al.}{2014}]{Torrey_2014}
{Torrey} P.,  {Vogelsberger} M.,  {Genel} S.,  {Sijacki} D.,  {Springel} V.,   {Hernquist} L.,  2014, \mn@doi [\mnras] {10.1093/mnras/stt2295}, \href {https://ui.adsabs.harvard.edu/abs/2014MNRAS.438.1985T} {438, 1985}

\bibitem[\protect\citeauthoryear{{Torrey} et~al.,}{{Torrey} et~al.}{2018}]{Torrey_2018}
{Torrey} P.,  et~al., 2018, \mn@doi [\mnras] {10.1093/mnrasl/sly031}, \href {https://ui.adsabs.harvard.edu/abs/2018MNRAS.477L..16T} {477, L16}

\bibitem[\protect\citeauthoryear{{Torrey} et~al.,}{{Torrey} et~al.}{2019}]{Torrey_2019}
{Torrey} P.,  et~al., 2019, \mn@doi [\mnras] {10.1093/mnras/stz243}, \href {https://ui.adsabs.harvard.edu/abs/2019MNRAS.484.5587T} {484, 5587}

\bibitem[\protect\citeauthoryear{{Tremonti} et~al.,}{{Tremonti} et~al.}{2004}]{Tremonti_2004}
{Tremonti} C.~A.,  et~al., 2004, \mn@doi [\apj] {10.1086/423264}, \href {https://ui.adsabs.harvard.edu/abs/2004ApJ...613..898T} {613, 898}

\bibitem[\protect\citeauthoryear{{Veilleux}, {Cecil}  \& {Bland-Hawthorn}}{{Veilleux} et~al.}{2005}]{Veilleux_2005}
{Veilleux} S.,  {Cecil} G.,   {Bland-Hawthorn} J.,  2005, \mn@doi [\araa] {10.1146/annurev.astro.43.072103.150610}, \href {https://ui.adsabs.harvard.edu/abs/2005ARA&A..43..769V} {43, 769}

\bibitem[\protect\citeauthoryear{{Veilleux}, {Maiolino}, {Bolatto}  \& {Aalto}}{{Veilleux} et~al.}{2020}]{Veilleux_2020}
{Veilleux} S.,  {Maiolino} R.,  {Bolatto} A.~D.,   {Aalto} S.,  2020, \mn@doi [\aapr] {10.1007/s00159-019-0121-9}, \href {https://ui.adsabs.harvard.edu/abs/2020A&ARv..28....2V} {28, 2}

\bibitem[\protect\citeauthoryear{{Vogelsberger}, {Sijacki}, {Kere{\v{s}}}, {Springel}  \& {Hernquist}}{{Vogelsberger} et~al.}{2012}]{Vogelsberger_2012}
{Vogelsberger} M.,  {Sijacki} D.,  {Kere{\v{s}}} D.,  {Springel} V.,   {Hernquist} L.,  2012, \mn@doi [\mnras] {10.1111/j.1365-2966.2012.21590.x}, \href {https://ui.adsabs.harvard.edu/abs/2012MNRAS.425.3024V} {425, 3024}

\bibitem[\protect\citeauthoryear{{Vogelsberger}, {Genel}, {Sijacki}, {Torrey}, {Springel}  \& {Hernquist}}{{Vogelsberger} et~al.}{2013}]{Vogelsberger_2013}
{Vogelsberger} M.,  {Genel} S.,  {Sijacki} D.,  {Torrey} P.,  {Springel} V.,   {Hernquist} L.,  2013, \mn@doi [\mnras] {10.1093/mnras/stt1789}, \href {https://ui.adsabs.harvard.edu/abs/2013MNRAS.436.3031V} {436, 3031}

\bibitem[\protect\citeauthoryear{{Vogelsberger} et~al.,}{{Vogelsberger} et~al.}{2014a}]{Vogelsberger_2014a}
{Vogelsberger} M.,  et~al., 2014a, \mn@doi [\mnras] {10.1093/mnras/stu1536}, \href {https://ui.adsabs.harvard.edu/abs/2014MNRAS.444.1518V} {444, 1518}

\bibitem[\protect\citeauthoryear{{Vogelsberger} et~al.,}{{Vogelsberger} et~al.}{2014b}]{Vogelsberger_2014b}
{Vogelsberger} M.,  et~al., 2014b, \mn@doi [\nat] {10.1038/nature13316}, \href {https://ui.adsabs.harvard.edu/abs/2014Natur.509..177V} {509, 177}

\bibitem[\protect\citeauthoryear{Weinberger et~al.,}{Weinberger et~al.}{2017}]{Weinberger_2017}
Weinberger R.,  et~al., 2017, \mn@doi [MNRAS] {10.1093/mnras/stw2944}, 465, 3291

\bibitem[\protect\citeauthoryear{{Wiersma}, {Schaye}, {Theuns}, {Dalla Vecchia}  \& {Tornatore}}{{Wiersma} et~al.}{2009}]{Wiersma_2009b}
{Wiersma} R. P.~C.,  {Schaye} J.,  {Theuns} T.,  {Dalla Vecchia} C.,   {Tornatore} L.,  2009, \mn@doi [\mnras] {10.1111/j.1365-2966.2009.15331.x}, \href {https://ui.adsabs.harvard.edu/abs/2009MNRAS.399..574W} {399, 574}

\bibitem[\protect\citeauthoryear{{Yang}, {Scholte}  \& {Saintonge}}{{Yang} et~al.}{2022}]{Yang_2022}
{Yang} N.,  {Scholte} D.,   {Saintonge} A.,  2022, \mn@doi [arXiv e-prints] {10.48550/arXiv.2212.10657}, \href {https://ui.adsabs.harvard.edu/abs/2022arXiv221210657Y} {p. arXiv:2212.10657}

\bibitem[\protect\citeauthoryear{{Yates}, {Kauffmann}  \& {Guo}}{{Yates} et~al.}{2012}]{Yates_2012}
{Yates} R.~M.,  {Kauffmann} G.,   {Guo} Q.,  2012, \mn@doi [\mnras] {10.1111/j.1365-2966.2012.20595.x}, \href {https://ui.adsabs.harvard.edu/abs/2012MNRAS.422..215Y} {422, 215}

\bibitem[\protect\citeauthoryear{{Yates}, {Henriques}, {Fu}, {Kauffmann}, {Thomas}, {Guo}, {White}  \& {Schady}}{{Yates} et~al.}{2021}]{Yates_2021}
{Yates} R.~M.,  {Henriques} B. M.~B.,  {Fu} J.,  {Kauffmann} G.,  {Thomas} P.~A.,  {Guo} Q.,  {White} S. D.~M.,   {Schady} P.,  2021, \mn@doi [\mnras] {10.1093/mnras/stab741}, \href {https://ui.adsabs.harvard.edu/abs/2021MNRAS.503.4474Y} {503, 4474}

\bibitem[\protect\citeauthoryear{{Zahid}, {Kewley}  \& {Bresolin}}{{Zahid} et~al.}{2011}]{Zahid_2011}
{Zahid} H.~J.,  {Kewley} L.~J.,   {Bresolin} F.,  2011, \mn@doi [\apj] {10.1088/0004-637X/730/2/137}, \href {https://ui.adsabs.harvard.edu/abs/2011ApJ...730..137Z} {730, 137}

\bibitem[\protect\citeauthoryear{{Zahid}, {Kudritzki}, {Conroy}, {Andrews}  \& {Ho}}{{Zahid} et~al.}{2017}]{Zahid_2017}
{Zahid} H.~J.,  {Kudritzki} R.-P.,  {Conroy} C.,  {Andrews} B.,   {Ho} I.~T.,  2017, \mn@doi [\apj] {10.3847/1538-4357/aa88ae}, \href {https://ui.adsabs.harvard.edu/abs/2017ApJ...847...18Z} {847, 18}

\bibitem[\protect\citeauthoryear{{van Loon}, {Mitchell}  \& {Schaye}}{{van Loon} et~al.}{2021}]{vanLoon_2021}
{van Loon} M.~L.,  {Mitchell} P.~D.,   {Schaye} J.,  2021, \mn@doi [\mnras] {10.1093/mnras/stab1254}, \href {https://ui.adsabs.harvard.edu/abs/2021MNRAS.504.4817V} {504, 4817}

\makeatother
\end{thebibliography}

%%%%%%%%%%%%%%%%%%%%%%%%%%%%%%%%%%%%%%%%%%%%%%%%%%

%%%%%%%%%%%%%%%%% APPENDICES %%%%%%%%%%%%%%%%%%%%%

\appendix

\section{Redshift evolution in TNG}
\label{appendix:TNGz=0}

\begin{figure*}
    \centering
    \includegraphics[width=\linewidth]{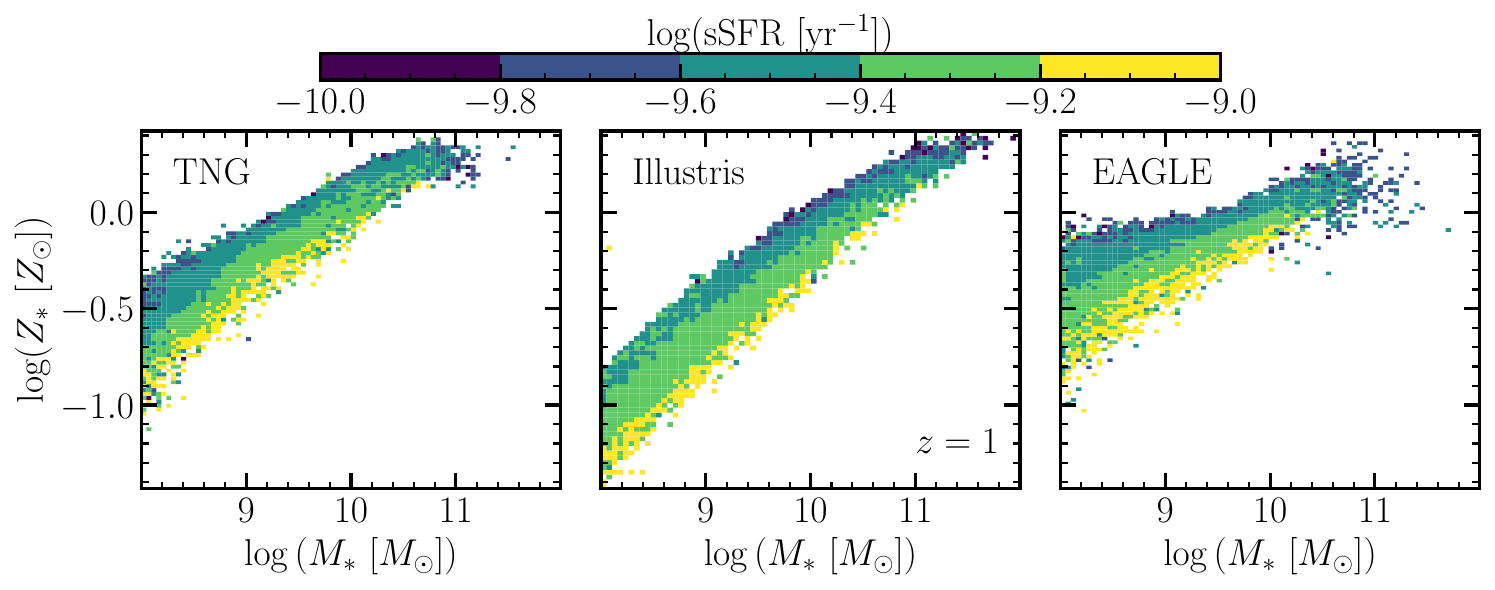}
    \caption{Same as Figure~\ref{fig:MZSFR*}, but at $z=1$. Note the clear evidence of a $M_*-Z_*-{\rm SFR}$ relation at lower masses  in the left panel for TNG.}
    \label{fig:AppendixVersion}
\end{figure*}

\begin{figure*}
    \centering
    \includegraphics[width=\linewidth]{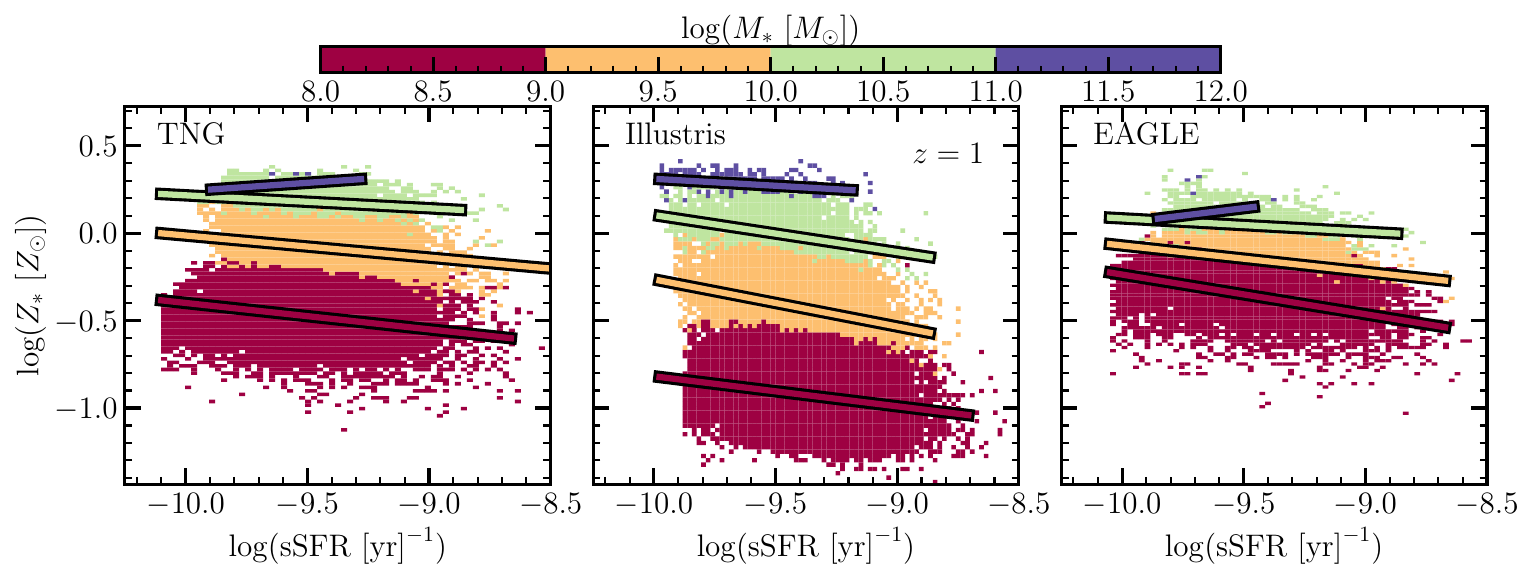}
    \caption{Same as Figure~\ref{fig:MZSFR_sSFR}, but at $z=1$. Note the clear evidence of a $M_*-Z_*-{\rm SFR}$ relation at lower masses in the left panel for TNG.}
    \label{fig:AppendixFig4}
\end{figure*}

As mentioned in Section~\ref{subsubsec:leastScatter} (and shown in Figures~\ref{fig:MZSFR*}~and~\ref{fig:alpha_redshift}), at redshift 0 in TNG, there is \edit{virtually} no tertiary dependence on SFR within the \MZsR{} (i.e., \edit{$\alpha\sim0.0$}).
Whereas in Figure~\ref{fig:MZSFR*} there is no clear trend of decreasing SFR with increasing metallicity in galaxies of stellar mass $\log (M_*~[M_\odot]) \lesssim 9.5$ in TNG, the left-most panel of Figure~\ref{fig:AppendixVersion} shows a much clearer trend with galaxies' SFRs.
Furthermore, the left panel of Figure~\ref{fig:AppendixFig4} shows that we recover (mostly\footnote{As mentioned in Section~\ref{subsec:residualCorrelations}, the highest mass bins have flatter (or even inverted) relationships; similar to the findings of \cite{Yates_2012}.}) negative slopes within the relationship between the sSFR and stellar metallicity at fixed stellar mass.

One possible source of the lack of a clear relationship in TNG at $z=0$ is the different implementations of galactic scale winds in the updated TNG models \citep[see][for a complete description of the differences between the models]{Pillepich_2018a}.
The original Illustris framework launches the winds as velocities dependent on the local dark matter velocity dispersion.
TNG takes this one step further and introduces a redshift dependence to the winds by setting a velocity floor for wind injection.
This velocity floor ensures that low mass haloes do not have unphysical mass loading factors.
Consequently, the low redshift star formation suppression and high redshift feedback are more effective in TNG compared to Illustirs.
These different wind implementations give rise to different \thirdedit{mass-metallicity} relations that evolve differently with time.
It is possible that this increased star formation suppression at low redshifts specifically plays a part in the lack of an $M_*-Z_*-{\rm SFR}$ relation at $z=0$ in TNG.

\begin{figure*}
    \centering
    \includegraphics[width=\linewidth]{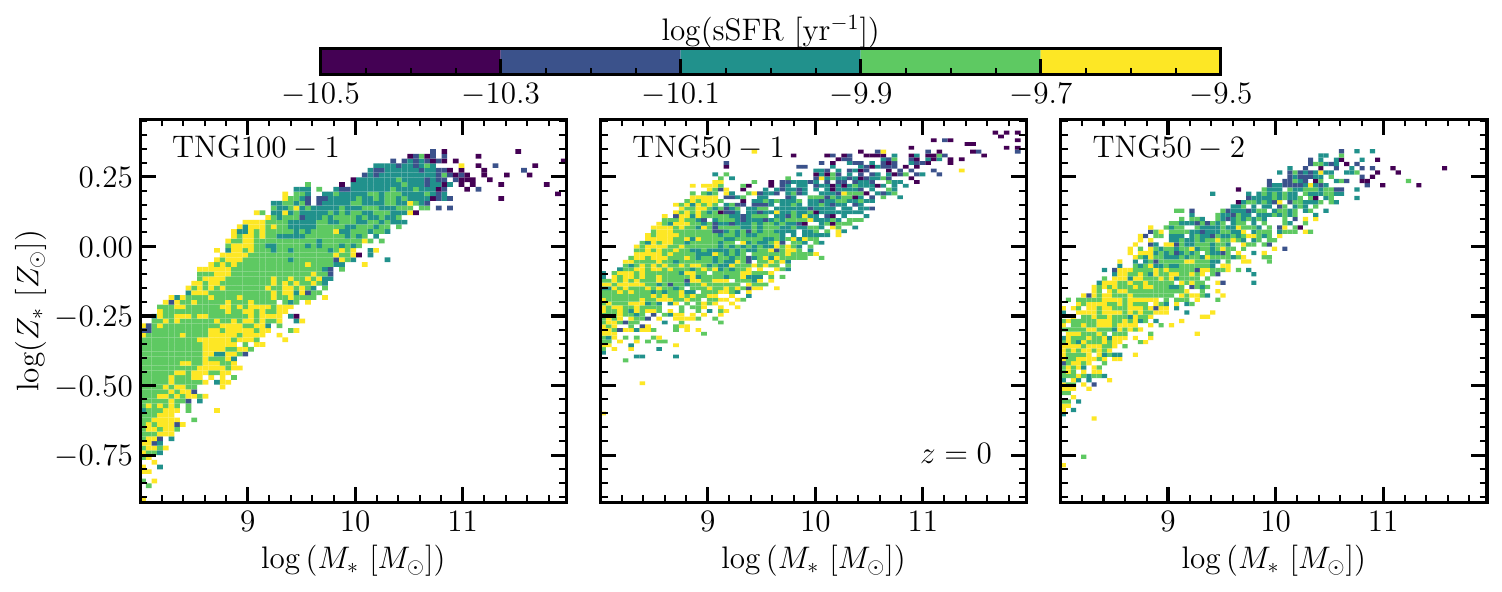}
    \includegraphics[width=\linewidth]{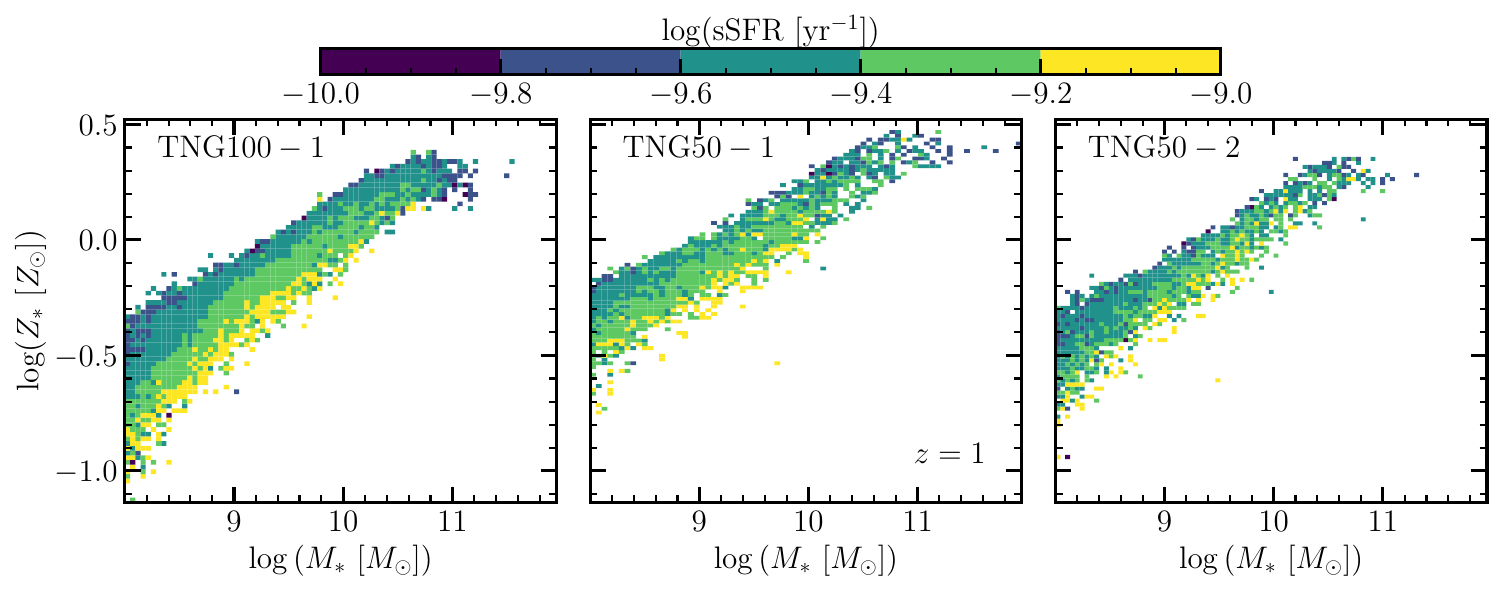}
    \caption{\secondedit{Same as Figure~\ref{fig:MZSFR*} for (left to right) TNG100-1 (called TNG in previous plots), TNG50-1, and TNG50-2. The top panels are at $z=0$ and the bottom panels are at $z=1$.}}
    \label{fig:50_comparison}
\end{figure*}

\secondedit{
This idea is further corroborated by investigating TNG50, a (51.7 Mpc)$^3$ box run of TNG \citep[][]{Pillepich_2019}.
Following all of the same selection criteria from Section~\ref{subsec:selectionCriteria}, we perform a similar analysis in TNG50-1 and TNG50-2, the highest resolution runs of TNG50.
TNG50-1 has an initial baryon mass resolution of $8.4\times10^4~M_\odot$ which is much higher resolution than TNG100-1.
TNG50-2's initial baryon mass resolution is more comparable to that of TNG100-1 at $6.7\times10^5~M_\odot$. 
We find a similar lack of dependence on sSFR at $z=0$ in both TNG50-1 and 50-2 (top panels of Figure~\ref{fig:50_comparison}).
The sSFR dependence, however, appears at $z=1$ in TNG50-1 and 50-2 (bottom panels of Figure~\ref{fig:50_comparison}).
This suggests that the \MZsR's lack of a dependence on sSFR at $z=0$ is likely a feature of the model and not a by-product of the mass resolution of TNG100-1.
}

It is interesting to note that a similar lack of correlation can be seen in \citeauthor{Torrey_2019} (\citeyear{Torrey_2019}; top left panel of their Figure 7) in the gas-phase metallicity.

\section{Dependence on the specific star formation main sequence selection criteria}
\label{appendix:sSFMS_dependence}

\begin{figure*}
    \centering
    \includegraphics[width=0.49\linewidth]{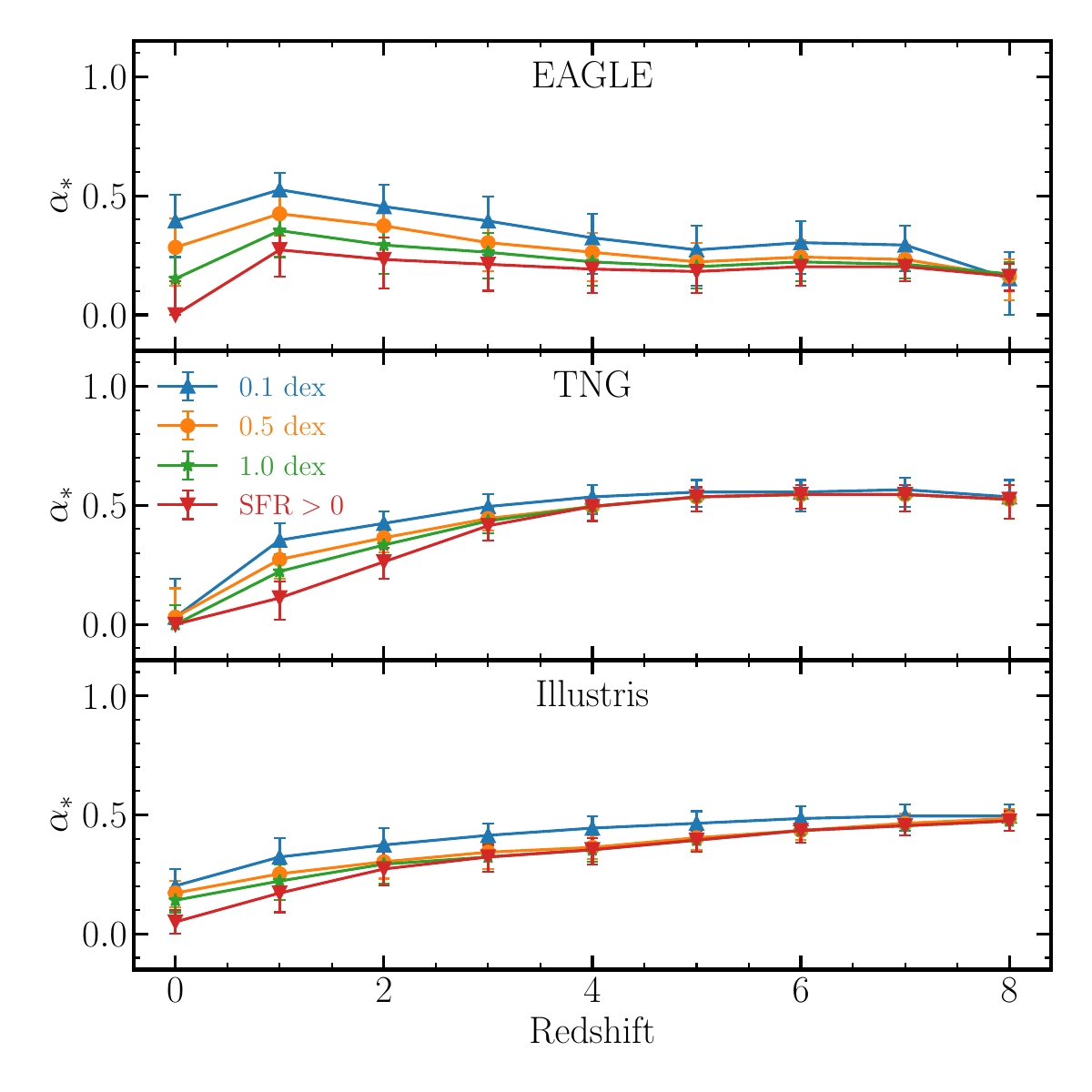}
    \includegraphics[width=0.49\linewidth]{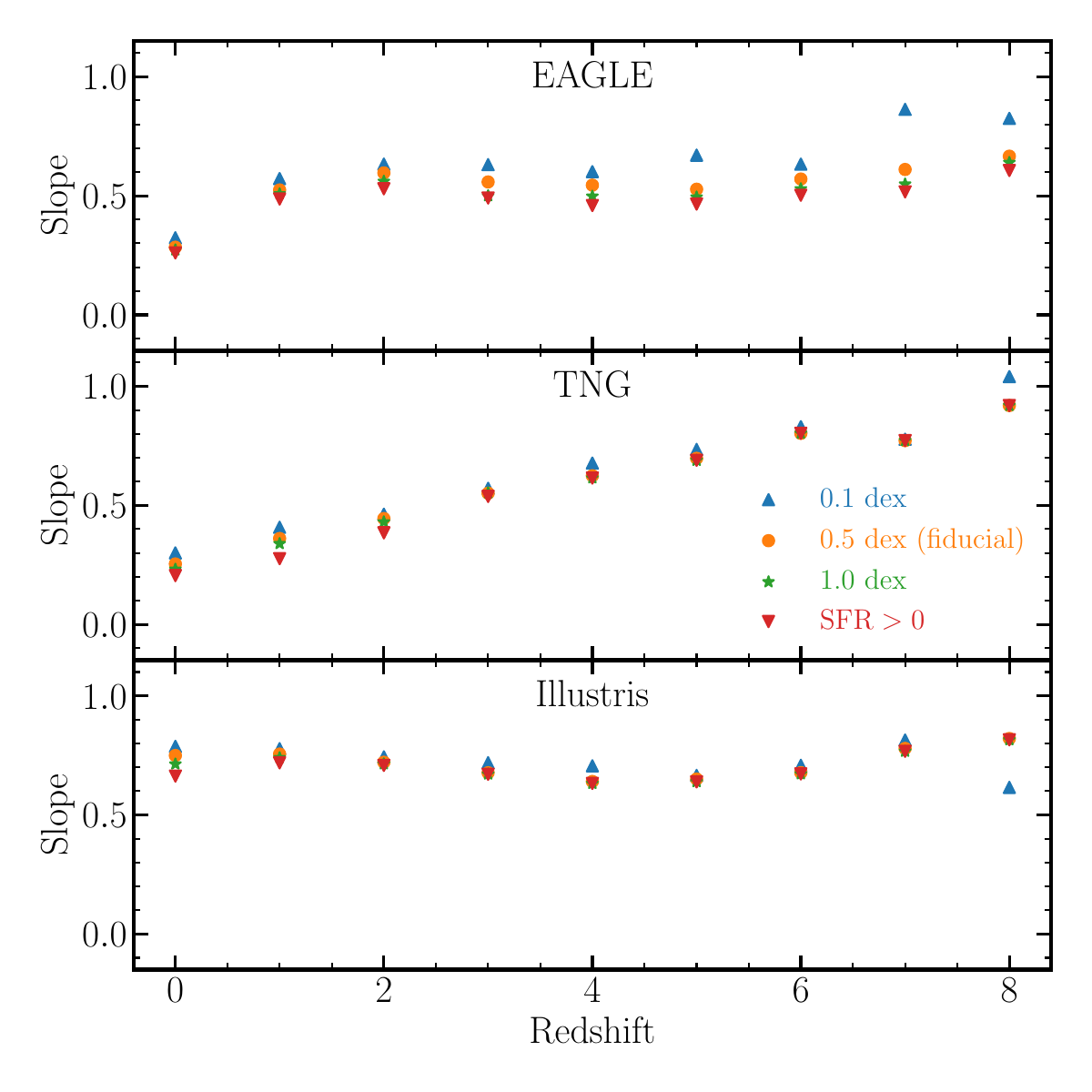}
    \caption{{\bf Left:} Determination of $\alpha$, the free parameter in Equation~\ref{eqn:mu}, for EAGLE, TNG, and Illustris (top, middle, bottom, respectively) as a function of redshift for the four sSFMS selection variations. {\bf Right:} Value for the slope of the offset relations (as in Figure~\ref{fig:offset_relations}) for EAGLE, TNG, and Illustris (top, middle, bottom, respectively) as a function of redshift for the four sSFMS selection variations.}
    \label{fig:appendixAlpha}
\end{figure*}

\edit{
When selecting galaxies for our sample, we restrict ourselves to only star forming galaxies characterised by a sSFMS (see Section~\ref{subsec:selectionCriteria}).
One of the key characteristics of the selection is that any galaxy that is significantly below the relation is classified as quiescent and omitted from the sample.
% In this appendix, to investigate the impact of that omission threshold on key results, we vary that threshold.
The fiducial classification is that any galaxy that is 0.5 dex below the sSFMS is omitted, in this appendix we consider three alterations: (i) any galaxy greater than 0.1 dex below the sSFMS is omitted, (ii) any galaxy greater than 1.0 dex below the sSFMS is omitted, and (iii) only galaxies not forming stars at all are omitted.

Using the each definition of the sSFMS, the left panels of Figure~\ref{fig:appendixAlpha} shows the determination of $\alpha$ (as in Figure~\ref{fig:alpha_redshift}, Section~\ref{subsubsec:leastScatter}) as a function of redshift in EAGLE, TNG, and Illustris in the top, middle, and bottom panels, respectively.
We find that across redshift $\alpha$ is relatively insensitive to changes in the sample, further the {\it quantitative} trends described in the main text are unchanged.
Similarly, the right panels of Figure~\ref{fig:appendixAlpha} shows the slopes of the offset relations between $\Delta Z_{\rm gas}$ and $\Delta Z_*$ (as in Figure~\ref{fig:slope_comparison}, Section~\ref{subsec:origin_MZR}) as a function of redshift for the the four different sample criteria in EAGLE, TNG, and Illustris in the top, middle, and bottom panels, respectively.
In this we also find that variations to the sample criteria produce only modest changes in the quantitative values of the slopes and the qualitative trends are largely preserved.
Thus, we conclude that our key results are fairly insensitive to the sSFMS criteria outlined in Section~\ref{subsec:selectionCriteria}.
}

% \section{Dependence on resolution}

% \input{tables/resolution}

% If you want to present additional material which would interrupt the flow of the main paper,
% it can be placed in an Appendix which appears after the list of references.

%%%%%%%%%%%%%%%%%%%%%%%%%%%%%%%%%%%%%%%%%%%%%%%%%%

% Don't change these lines
\bsp	% typesetting comment
\label{lastpage}
\end{document}